\documentclass[10pt,twocolumn,twoside,journal]{IEEEtran}
\IEEEoverridecommandlockouts
\usepackage{cite}
\usepackage{amsmath,amssymb,amsfonts}

\usepackage{algorithmic}
\usepackage{graphicx}
\usepackage{textcomp}
\usepackage{xcolor}

\ifCLASSOPTIONcompsoc
\usepackage[caption=false,font=normalsize,labelfon
t=sf,textfont=sf]{subfig}
\else
\usepackage[caption=false,font=footnotesize]{subfig}
\fi

\usepackage{multirow}
\usepackage{array}

\usepackage{lipsum}     
\usepackage{multicol}   
\usepackage{balance}
\DeclareMathOperator*{\argmax}{arg\,max}

\usepackage{booktabs}
\usepackage{algorithm}

\usepackage{dblfloatfix}

\makeatletter
\def\ps@IEEEtitlepagestyle{%
  \def\@oddfoot{\mycopyrightnotice}%
  \def\@oddhead{\hbox{}\@IEEEheaderstyle\leftmark\hfil\thepage}\relax
  \def\@evenhead{\@IEEEheaderstyle\thepage\hfil\leftmark\hbox{}}\relax
  \def\@evenfoot{}%
}
\def\mycopyrightnotice{%
  \begin{minipage}{\textwidth}
  \centering \scriptsize
  Copyright~\copyright~2023 IEEE. Personal use of this material is permitted. Permission from IEEE must be obtained for all other uses, in any current or future media, including reprinting/republishing this material for advertising or promotional purposes, creating new collective works, for resale or redistribution to servers or lists, or reuse of any copyrighted component of this work in other works.
  \end{minipage}
}
\makeatother

\def\BibTeX{{\rm B\kern-.05em{\sc i\kern-.025em b}\kern-.08em
    T\kern-.1667em\lower.7ex\hbox{E}\kern-.125emX}}
\begin{document}

\title{In-Situ Calibration of Antenna Arrays for Positioning With 5G Networks
}

\author{Mengguan~Pan,~\IEEEmembership{Member,~IEEE,}
  Shengheng~Liu,~\IEEEmembership{Senior Member,~IEEE,}
  Peng~Liu,
  Wangdong~Qi,~\IEEEmembership{Member,~IEEE,}
  Yongming~Huang,~\IEEEmembership{Senior Member,~IEEE,}
  Wang~Zheng,
  Qihui~Wu,~\IEEEmembership{Senior Member,~IEEE,}\\
  Markus Gardill,~\IEEEmembership{Member,~IEEE}
  \vspace{-1.6em}
  \thanks{Manuscript received 29 December 2022; revised 16 February 2023; accepted xx xxx 2023. Date of publication xx xxx 2023; date of current version xx xxx 2023. This research was supported in part by the National Natural Science Foundation of China under Grant Nos. 62001103 and U1936201. \textit{(Corresponding authors: Shengheng Liu and Peng Liu.)}}
        \thanks{Mengguan Pan and Wang Zheng are with the Purple Mountain Laboratories, Nanjing 211111, China (e-mail: panmengguan@outlook.com).}
        \thanks{Shengheng Liu, Wangdong Qi, and Yongming Huang are with the National Mobile Communications Research Laboratory, Southeast University, Nanjing 210096, China, and also with the Purple Mountain Laboratories, Nanjing 211111, China (e-mail: s.liu@seu.edu.cn, qiwangdong@pmlabs.com.cn, huangym@seu.edu.cn).}
        \thanks{Peng Liu is with the College of Electronic and Information Engineering, Nanjing University of Aeronautics and Astronautics, Nanjing 211106, China, and also with the Department of Network Engineering, Army Engineering University of PLA, Nanjing 210007, China (e-mail: herolp@gmail.com).}
        \thanks{Qihui Wu is with the College of Electronic and Information Engineering, Nanjing University of Aeronautics and Astronautics, Nanjing 211106, China.}
        \thanks{Markus Gardill is with Brandenburg University of Technology Cottbus--Senftenberg, Cottbus, 03046, Germany.}
      }

      \markboth{IEEE TRANSACTIONS ON MICROWAVE THEORY AND TECHNIQUES, ~Vol.~XX, No.~X, XXX~2023}%
      {PAN \MakeLowercase{\textit{et al.}}: In-Situ Calibration of Antenna Arrays for Positioning With 5G Networks}

\maketitle
\IEEEpubidadjcol

\newcommand{\figpath}{./figs/}

\begin{abstract}
  Owing to the ubiquity of cellular communication signals, positioning with the fifth generation (5G) signal has emerged as a promising solution in global navigation satellite system-denied areas.
  Unfortunately, although the widely employed antenna arrays in 5G remote radio units (RRUs) facilitate the measurement of the direction of arrival (DOA), DOA-based positioning performance is severely degraded by array errors.
  This paper proposes an in-situ calibration framework with a user terminal transmitting 5G reference signals at several known positions in the actual operating environment and the accessible RRUs estimating their array errors from these reference signals.
  Further, since sub-6GHz small-cell RRUs deployed for indoor coverage generally have small-aperture antenna arrays, while 5G signals have plentiful bandwidth resources, this work segregates the multipath components
  via super-resolution delay estimation based on the maximum likelihood criteria.
  This differs significantly from existing in-situ calibration works which
  resolve multipaths in the spatial domain.
  The superiority of the proposed method is first verified by numerical simulations.
  We then demonstrate via field test with commercial 5G equipment that, a reduction of 46.7\% for \(1{\text -}\sigma\) DOA estimation error can be achieved by in-situ calibration using the proposed method.
\end{abstract}

\begin{IEEEkeywords}
  5G positioning, angle-of-arrival (AOA), array calibration, direction-of-arrival (DOA), field test, in-situ calibration, multipath, wireless localization.
\end{IEEEkeywords}

\section{Introduction}
\IEEEPARstart{P}{recise} positioning is the key enabler for a wide range of emerging applications such as indoor navigation \cite{el-sheimy2021_IndoorNavigationSt}, autonomous driving \cite{kuutti2018_SurveyStateoftheArt}, healthcare \cite{paolini2019_FallDetection3D}, intelligent transportation \cite{dobrev2017_SteadyDeliveryWire}, industrial internet of things \cite{lohan2018_BenefitsPositioning}, etc.
With the progressive deployment of the 5G small-cell (i.e. microcell, picocell, or femtocell) base stations (a.k.a. gNodeBs, or gNBs) in GNSS challenging scenarios \cite{rodriguez2014_SmallCells5G}, such as deep urban canyons, tunnels, undergrounds, or indoor environments, the abundant 5G signals become a promising candidate for achieving accurate and reliable positioning in these areas \cite{kassas2017_HearThereforeKnow, delperal-rosado22_SurveyCellularMobi, dwivedi2021_Positioning5GNetwo}.

The widespread employment of antenna array technique for small-cell 5G RRUs
has attracted growing interest in exploiting the DOA information for 5G
positioning \cite{sun2021_ComparativeStudy3D, koivisto2017_JointDevicePositio,
  menta2019_PerformanceAoABase}, as it not only obviates the need for precise
timing-synchronizations but is also an indispensable measurement for
implementing single site positioning \cite{sun2021_ComparativeStudy3D}.
However, the DOA estimation performance is inevitably impaired by the nonideal
responses of the antenna arrays and the RF channels of the receiver
\cite{swindlehurst1992_PerformanceAnalysis,
  swindlehurst1993_PerformanceAnalysis}. The majority of existing works on
DOA-based wireless positioning either assume an ideal array model
\cite{koivisto2017_JointDevicePositio, sun2021_ComparativeStudy3D} or merely
take into account the gain-phase errors introduced by the RF channels
\cite{xiong2013_ArrayTrackFinegrain, kotaru2015_SpotFiDecimeterLeva,
  shamaei2021_JointTOADOAa}. However, the antenna array per se is also
suffered from severe imperfections (a.k.a. array errors), resulting in the
deviations of the real array manifold from the theoretic one. Therefore,
precise array calibration, which amounts to estimating the array error and
deriving the real array manifold, is pivotal to achieving high-accuracy
DOA-based positioning.

According to the adopted model for array errors,
  calibration can be achieved by using either parametric methods or
  non-parametric methods. Parametric methods only consider typical
  array errors, i.e. gain-phase errors, mutual couplings, and element
  location perturbations, and model them with a small number of
  direction-independent parameters \cite{liu2013_UnifiedFrameworkSp,
    wang2019_DOAEstimationMutua, chen2020_NewAtomicNorma,
    pierre1991_ExperimentalPerform, song2021_MaximumLikelihoodSa,
    yamada2011_ArrayCalibrationTe, groschel2017_SystemConceptOnlin,
    pohlmann2022_BayesianInSituCali, sippel2020_InSituCalibrationA,
    gupta2003_ExperimentalStudyA}. However, apart from these common
  array errors, real-world antennas are generally impaired by other
  unpredictable and more complicated imperfections, such as the
  electromagnetic interactions between the array and nearby
  structures, manufacturing inaccuracy, etc., which can hardly be
  captured by the parametric models, as verified by field experiments
  in \cite{gupta2003_ExperimentalStudyA}. On the contrary,
  non-parametric methods, which gather all the array nonidealities
  into a direction-dependent \cite{friedlander2018_AntennaArrayManifo}
  (also known as scan-dependent in
  \cite{lanne2006_CalibratingArraySc}) error function
  \cite{see1995_MethodArrayCalibra, lanne2006_CalibratingArraySc,
    heidenreich2009_HighresolutionDirec,
    ibanezurzaiz2021_DigitalBeamforming,
    leshem2000_ArrayCalibrationPr, pan2021_IndoorDirectPositi,
    friedlander2018_AntennaArrayManifo,
    vasanelli2020_CalibrationDirectio, yang2022_TheoryExperimentAr,
    pan2022_EfficientJointDOAb}, can depict arbitrary array error
  patterns.

Calibration techniques can also be categorized as chamber calibration, in-situ calibration, and self-calibration methods.
Measuring the array response in an anechoic chamber is the standard
way for array calibration \cite{vasanelli2020_CalibrationDirectio,
  pan2022_EfficientJointDOAb, yang2022_TheoryExperimentAr}.
However, the array manifold in its working environments is hardly the same as the nominal manifold measured in the anechoic chamber, owing to
installation errors, scatterings from array mounting structure and nearby objects, coupling behavior changes, etc \cite{gupta2003_ExperimentalStudyA, sippel2020_InSituCalibrationA}. Also, calibrating every antenna array in an anechoic chamber and reinstalling them in their working positions is costly and time-consuming.

Self-calibration
circumvents the aforementioned drawbacks of chamber calibration by estimating simultaneously both the wavefield DOA and the array errors based on online measurements \cite{liu2013_UnifiedFrameworkSp, chen2020_NewAtomicNorma, wang2019_DOAEstimationMutua}.
It commonly employs the parametric array error model to reduce the number of unknowns and optimizes a cost function for joint estimation.
However, the global optimality of this multi-dimensional estimation problem is not always guaranteed and ambiguities arise for some array geometries \cite{pierre1991_ExperimentalPerform, hung1994_CriticalStudySelfc}.
Moreover, it also suffers from high computational complexity owing to the vast parameter space.

In-situ calibration measures the array error in the actual operating environment using auxiliary calibration sources whose positions, which are also denoted as the CPPs, are known \cite{pierre1991_ExperimentalPerform, see1995_MethodArrayCalibra, leshem2000_ArrayCalibrationPr, yamada2011_ArrayCalibrationTe, pan2021_IndoorDirectPositi, sippel2020_InSituCalibrationA, song2021_MaximumLikelihoodSa, heidenreich2009_HighresolutionDirec, ibanezurzaiz2021_DigitalBeamforming, pohlmann2022_BayesianInSituCali, gupta2003_ExperimentalStudyA, groschel2017_SystemConceptOnlin, lanne2006_CalibratingArraySc}.
It reaches a reasonable compromise between chamber calibration and self-calibration as it obtains the exact in-field array response with significantly lower complexity and much better accuracy than self-calibration.

However, the in-field calibration signal is inevitably interfered by the multipath effect caused by reflections or scatterings of surrounding objects.
The most popular solution in current literature is to approximate the array steering vector with the PE of the spatial covariance matrix of the received signal \cite{see1995_MethodArrayCalibra, lanne2006_CalibratingArraySc, heidenreich2009_HighresolutionDirec, ibanezurzaiz2021_DigitalBeamforming}.
Its performance is guaranteed only when the LOS signal power dominates all the NLOS propagated signal powers.
Besides, the coherency between these multipath components (including both LOS and NLOS paths) also deteriorates the approximation performance.
Leshem and Wax \cite{leshem2000_ArrayCalibrationPr} and Yamada \textit{et al.} \cite{yamada2011_ArrayCalibrationTe} explicitly consider the multipaths in the signal model for in-situ calibration. However, the former work \cite{leshem2000_ArrayCalibrationPr} depends on a complex calibration scheme that includes physically rotating the array and transmitting calibration signals in two different locations.
The latter work \cite{yamada2011_ArrayCalibrationTe} assumes a parametric array error model with only gain-phase errors and mutual couplings.
Moreover, all the aforementioned in-situ calibration methods
work on the premise that the multipaths are separable by the spatial resolution of the antenna array.

Besides, other research efforts in counteracting the multipath effect for in-situ calibration include: Pan \textit{et al.} \cite{pan2021_IndoorDirectPositi}, who solve it from the statistical perspective by treating the summation of the multipath components as Gaussian noise based on the Rayleigh fading assumption; and Sippel \textit{et al.} \cite{sippel2020_InSituCalibrationA}, who propose to choose the CPPs within the array's near field to elude multipath effects.

As clearly demonstrated in our prior work
  \cite{pan2022_EfficientJointDOAb} with real-measured data from 5G
  sub-6GHz picocell RRUs, the array errors of a small-aperture antenna
  array exhibit noticeable dependency on incident directions. This
  arbitrary direction-dependent pattern can hardly be decomposed into
  parametric models. On the other hand, as discussed above, although
  some solutions exist for multipath mitigation in array calibration
  literature, they rely entirely on the array's spatial aperture to
  resolve multipath components and are inapplicable to small-scale
  arrays equipped by the small-cell 5G RRUs. Therefore, this paper
  attempts to solve the in-situ array calibration problem with
  explicit modelings of both the direction-dependent array errors and
  the multipath effects, and aims at providing a universal and pragmatic
  solution to in-situ calibrate the pervasively established 5G
  small-cell RRUs to support accurate DOA estimation and positioning.
Specifically, the main contributions of this work are summarized as
follows.
\begin{enumerate}
\item We design a comprehensive scheme for accurate array calibration
  by non-parametrically modeling the
    direction-dependent array errors and explicitly considering the
    multipath effects. Existing works either consider the
    non-parametric array model in an ideal environment or tackle the
    multipaths with a simplified parametric model assumed.  
\item
  We propose an in-situ array calibration framework that is easily deployable on existing commodity 5G infrastructure by exploiting the standard 5G reference signal as the calibration source and using the ready-to-use baseband channel estimates for array manifold estimation. As a result, it obviates any modification to the hardware or protocol and is scalable to calibrate these pervasively installed 5G RRUs. 
\item
  For small-cell RRUs whose apertures are
    extremely small, we propose to segregate the multipaths by their
    TOAs in estimating the array manifold, which is superior to the
    conventional way of resolving them in the spatial (angular)
    domain. In this vein, a joint array response and TOA estimation
  problem is formulated using the maximum likelihood criterion and
  solved via the computationally efficient EM
  approach. To the best of our knowledge, this is the first attempt to
  leverage signal bandwidth for multipath resolution in in-situ
  calibration.
\item We prototype a 5G positioning system with commercial 5G picocell RRUs
  and conduct extensive indoor field tests in an area of nearly
  \(1125\;\mathrm{m}^2\). We demonstrate: (i) the disparity between the
  in-situ and the nominal array manifold, and (ii) a reduction of \(46.7\%\)
  for \(1{\text -}\sigma\) DOA estimation error achieved by calibrating with
  the proposed method.
\end{enumerate}

The rest of this paper is organized as follows. Section \ref{sec:framework} presents the in-situ array calibration framework for 5G RRUs and describes the signal model for the array manifold estimation. Then the algorithm for estimating the array manifold in-situ is proposed in Section \ref{sec:array_error_estimation}. Next in Section \ref{sec:simulation} and Section \ref{sec:field_tests}, the calibration performance of the proposed method in multipath environments is evaluated by numerical simulations and field tests, respectively. 
Finally, Section \ref{sec:conclusion} concludes the paper. 

Notations: Boldface lowercase and uppercase letters respectively denote vectors and matrices, where vectors are by default in column orientation. Italic English letters and lowercase Greek letters denote scalars. Blackboard-bold characters denote number sets, in particular, \(\mathbb{R}\) and \(\mathbb{C}\) represent the sets of real and complex numbers, respectively. For convenience, remaining notations and abbreviations used in this article are explained in the Nomenclature Section.
 
\section{In-Situ Calibration Framework and Signal Model}
\label{sec:framework}
\subsection{In-Situ Calibration Framework}
\label{sec:insitu_framework}
\begin{figure}[htb]
  \centering
  \includegraphics[width=0.49\textwidth]{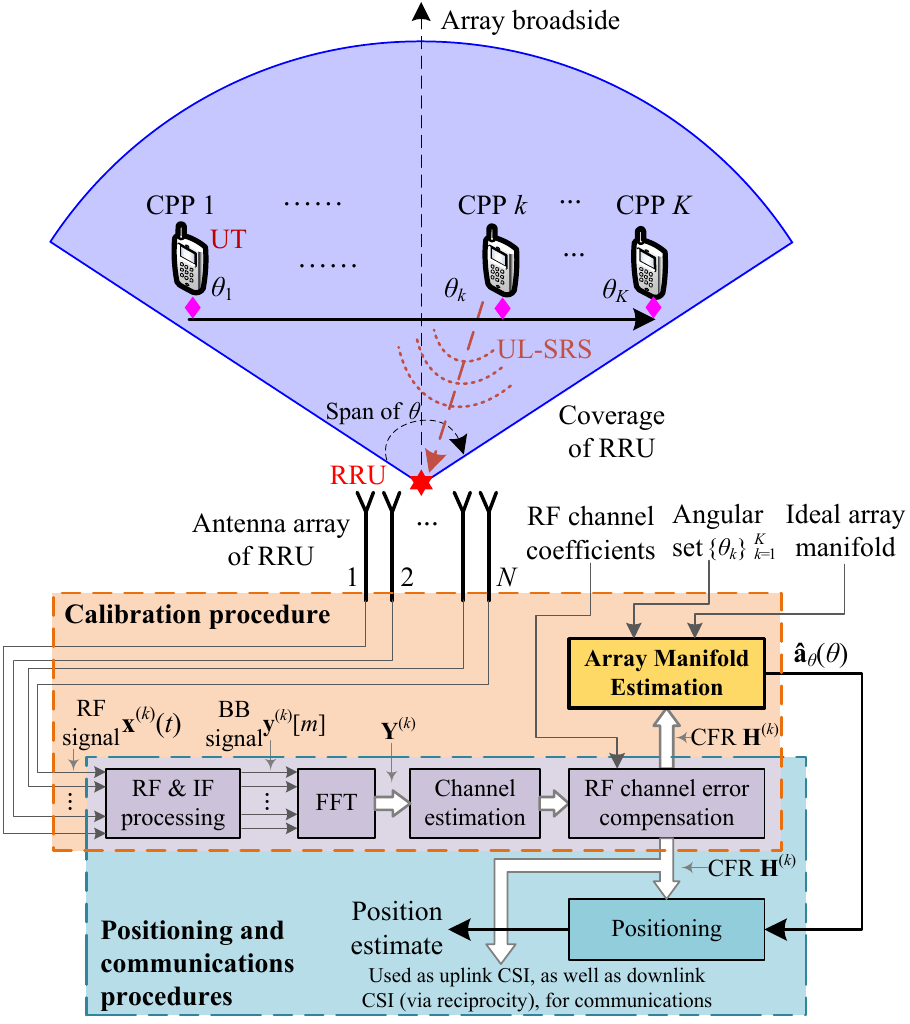}
  \caption{Illustration of measurement setup for in-situ calibration of array errors for 5G RRU.}
  \vspace{-1em}
  \label{fig:measurement_setup}
\end{figure}

An in-situ array error calibration framework is proposed for 5G RRU and is illustrated in Fig. \ref{fig:measurement_setup}.
All the measurement procedures shown in Fig. \ref{fig:measurement_setup} are conducted in a real working environment with a gNB installed.
The UT is placed at \(K\) known positions (CPPs) at which it sends the UL-SRS to the RRU.
CFRs perceived from these UL-SRSs, i.e \(\{\mathbf{H}^{(k)}\}_{k=1}^K\) in Fig. \ref{fig:measurement_setup}, are the data source for array calibration.
To calibrate the direction-dependent errors for signals impinged from all possible directions, the CPPs should be distributed over the entire coverage area.
Further, it is noticeable from Fig. \ref{fig:measurement_setup} that the calibration procedure shares the identical transmit waveform and front-end signal processing chain with the standard 5G communications and positioning procedures \cite{dwivedi2021_Positioning5GNetwo, 3gpp.38.857}.
This indicates that the proposed in-situ calibration framework not only eludes extra hardware and protocol overheads, but also precisely delineates all the hardware impairments suffered by the positioning signals.

It is worthwhile emphasizing that, this paper mainly considers the imperfect array response induced by the antennas, while that induced by the RF channels is compensated with pre-measured RF channel coefficients in both calibration and positioning procedures, as demonstrated in Fig. \ref{fig:measurement_setup}. The RF channel calibration coefficients are commonly measured by the means of internal calibration \cite{guo2020_MillimeterWave3DIm}. For small-cell RRUs whose transceivers lack internal calibration circuits, such as the dedicated calibration channel and the calibration network \cite{dreher2003_AntennaReceiverSys}, the RF channel coefficients can be measured by directly conducting the calibration signal from a 5G test UT or a signal generator to the receiving RF channels of the RRU via coaxial cables and an RF power splitter \cite{pan2022_EfficientJointDOAb}.

\subsection{Signal Model}
As shown in Fig. \ref{fig:measurement_setup}, after the RF and IF processing, the wireless channel response is estimated from the BB UL-SRS.  
Assuming that a UL-SRS with \(M\) subcarriers is transmitted at the frequency of \(f^{(c)}\) and impinges on an RRU via \(L\) propagation paths (i.e., \(L\) multipath components), then the CFR sensed by an \(N\)-element antenna array can be represented as \cite{pan2022_EfficientJointDOAb}
\begin{equation}
  \mathbf{H} = \sum\limits_{l = 1}^L \left[{{{\tilde \gamma }_l} \cdot {{\mathbf{a}}_\tau }({{\tilde \tau }_l}){\mathbf{a}}_\theta^{\mathsf{T}}({{\tilde \theta }_l})}\right] + \mathbf{W},
  \label{eq:Hmatrix}
\end{equation}
where \(\tilde{\theta}_l\), \(\tilde{\tau}_l\), and \(\tilde{\gamma}_l\) are the incident direction (a.k.a. DOA), propagation delay (a.k.a. TOA), and complex gain of the \(l\)-th path.
We assume a LOS scenario for in-situ calibration, hence the existence of \(l_{\mathrm{LOS}}\in\{1,2,\dots,L\}\) which indicates the LOS path index.
\(\mathbf{H}\in\mathbb{C}^{M\times N}\) and \(\mathbf{W}\in\mathbb{C}^{M\times N}\) in \eqref{eq:Hmatrix} represent the CFR matrix and the noise, respectively.
Entries of \(\mathbf{W}\) are i.i.d. complex-valued Gaussian noise with zero-mean and variance \(\sigma_w^2\), i.e. \(\left[\mathbf{W}\right]_{m,n}\sim\mathcal{CN}\left(0, \sigma_w^2\right)\).

Further, \(\mathbf{a}_\tau(\cdot)\) is the delay signature function whose value for the input delay of \(\tau\) is
\begin{equation}
  \mathbf{a}_\tau(\tau) = \exp\left(-\jmath2\pi\left[f_1,\dots,f_M\right]^{\mathsf{T}}\tau\right),
\end{equation}
where \(f_m = f^{(c)} + (m - \frac{M}{2})\Delta f\), \(\Delta f\) is the subcarrier spacing.
\(\mathbf{a}_\theta(\cdot)\) denotes the real array manifold which is actually a function of the continuous DOA.
Its value at a specific DOA of \(\theta\) is also known as the steering vector which, according to the direction-dependent array error model, is represented by
\begin{equation}
  \label{eq:direction-dependent}
\mathbf{a}_\theta(\theta) = \mathbf{a}_\theta'(\theta)\odot\boldsymbol{\zeta}(\theta),
\end{equation}
in which \(\boldsymbol{\zeta}(\cdot): \mathbb{U}\to\mathbb{C}^{N\times1}\) denotes the array modeling error function, where \(\mathbb{U}\subseteq\mathbb{R}\) is the interested range of DOA.
The \(n\)-th element of \(\boldsymbol{\zeta}(\theta)\) is \(\left[\boldsymbol{\zeta}(\theta)\right]_n = g_n(\theta)\cdot\exp\left(\jmath\varphi_n(\theta)\right)\), where \(g_n(\theta)\) and \(\varphi_n(\theta)\) represent the array gain and phase errors suffered by the \(n\)-th element for signals from the direction of \(\theta\).
\(\mathbf{a}_\theta'(\cdot)\) is the ideal array manifold and for a linear array, the value of \(\mathbf{a}_\theta'(\theta)\) is given by
\begin{equation}
  \mathbf{a}_\theta'(\theta) = \exp\left(\jmath\frac{2\pi}{\lambda}\left[d_1,\dots,d_N\right]^{\mathsf{T}}\sin\theta\right),
\end{equation}
where \(d_n\) is the position for the \(n\)-th antenna element and \(\lambda\) is the wavelength\footnote{Note that the received CFR signal model of \eqref{eq:Hmatrix} has separated signature functions for delay and angular domains. This indicates that the far-field model is adopted throughout this paper. For wireless positioning with 5G small-cell RRUs operated in the sub-6GHz frequency band, this assumption is reasonable, as the corresponding Fraunhofer distance is usually below \(1\;\mathrm{m}\) \cite{selvan2017_FraunhoferFresnelD}.}.

Array calibration amounts to the process of estimating the array error function \(\boldsymbol{\zeta}(\cdot)\) and deriving the actual manifold \(\mathbf{a}_\theta(\cdot)\).
Since DOA is essentially obtained from the phase shifts between antenna elements for a far-field signal model, this paper only considers the phase errors. Then the crucial issue of array calibration is reduced to the estimation of phase error functions \(\varphi_n(\theta), n=1,\dots,N\) based on measured CFRs, which will be studied in Section \ref{sec:array_error_estimation}. 

\section{Array manifold estimation}
\label{sec:array_error_estimation}

This section presents the proposed in-situ array calibration algorithm, i.e. algorithm for array manifold estimation shown in the framework of Fig. \ref{fig:measurement_setup}. To give an insight, we derive the CFR model of each receiving channel from equation (\ref{eq:Hmatrix}) as follows:
\begin{equation}
  \mathbf{h}_n = \mathbf{A}_\tau(\boldsymbol{\tau})\boldsymbol{\xi}_n + \mathbf{w}_n,\quad n = 1, \dots,N,
  \label{eq:h_n}
\end{equation}
where \(\mathbf{h}_n\) and \(\mathbf{w}_n\) are the \(n\)-th columns of \(\mathbf{H}\) and \(\mathbf{W}\) respectively; \(\boldsymbol{\tau} = [\tilde{\tau}_1,\dots,\tilde{\tau}_L]^{\mathsf{T}}\), \(\boldsymbol{\xi}_n = [\xi_{n1},\dots,\xi_{nL}]^{\mathsf{T}}\), and \(\mathbf{A}_\tau(\boldsymbol{\tau}) = \left[\mathbf{a}_\tau(\tilde{\tau}_1),\dots,\mathbf{a}_\tau(\tilde{\tau}_L)\right]\) are the collections of path delays, path gains, and delay signature vectors of all paths respectively, in which \(\xi_{nl} = \tilde{\gamma}_l\cdot\left[\mathbf{a}_\theta(\tilde{\theta}_l)\right]_n\) denotes the complex gain of the \(l\)-th path observed by the \(n\)-th antenna.

\begin{figure}[htb]
  \centering
  \includegraphics[width=0.49\textwidth]{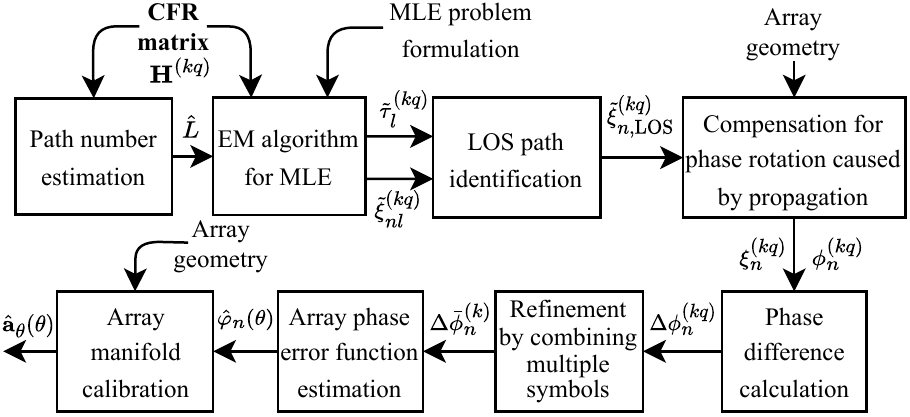}
  \caption{Flowchart for the proposed array manifold estimation algorithm.}
  \label{fig:calibration_flowchart}
\end{figure}

In the process of in-situ calibration, \(\mathbf{h}_n\) and \(\mathbf{a}_\tau(\cdot)\) are known while \(\boldsymbol{\xi}_n\) and \(\boldsymbol{\tau}\) are unknown parameters to be solved. The problem of joint estimating the path gain \(\boldsymbol{\xi}_n\) and path delay (TOA) \(\boldsymbol{\tau}\) is an inverse problem, i.e. the model parameters \(\boldsymbol{\xi}_n\) and \(\boldsymbol{\tau}\) that produce the observations \(\mathbf{h}_n\) need to be determined \cite{desmal2022_TrainedIterativeSh}. Further, since the complex gain induced by the wireless propagation (namely \(\tilde{\gamma}_l\)) is the same for all antenna elements, it can be easily canceled by retrieving the phase differences between \(\boldsymbol{\xi}_n,n=1,\dots,N\). Then the remaining term of \(\xi_{nl}\) is the array response \(\left[\mathbf{a}_\theta(\tilde{\theta}_l)\right]_n\). Therefore, this inverse problem is also denoted as the joint array response and TOA estimation problem.
Based on this idea, the detailed calibration procedure is designed as shown in Fig. \ref{fig:calibration_flowchart}.

First, to resolve and segregate the multipath components, an MLE problem is formulated. It is a multiple measurement vector problem with \(N\) ``spatial snapshots'' of the wireless channel and has \(2LN + L\) real unknown parameters, which can be gathered in a vector \(\boldsymbol{\Theta} = [\boldsymbol{\tau}^{\mathsf{T}}, \boldsymbol{\xi}_1^{\mathsf{T}}, \dots, \boldsymbol{\xi}_{N}^{\mathsf{T}}]^{\mathsf{T}}\).
Denoting the joint probability density of the CFR measurements as \(f(\mathbf{H};\boldsymbol{\Theta})\), which is also the likelihood function of parameters \(\boldsymbol{\Theta}\), 
then according to the signal model represented by equation (\ref{eq:h_n}) and the assumed i.i.d. Gaussian distribution of the noise components, the likelihood function is
\begin{equation}
  f(\mathbf{H};\boldsymbol{\Theta}) = \frac{1}{\left(\pi\sigma_w^2\right)^{MN}}\exp\left(-\frac{\sum\limits_{n=1}^N\left\|\mathbf{h}_n - \mathbf{A}(\boldsymbol{\tau})\boldsymbol{\xi}_n\right\|^2}{\sigma_w^2}\right).
\end{equation}
Taking the negative logarithm of this likelihood function and ignoring the terms that do not depend on any element in \(\boldsymbol{\Theta}\), the MLE problem for the joint estimation of parameters \(\boldsymbol{\Theta}\) is equivalent to
\begin{align}
  \label{eq:NLMLE}\hat{\boldsymbol{\Theta}} &= \min_{\boldsymbol{\Theta}} g(\mathbf{H}; \boldsymbol{\Theta}), \\
  \label{eq:opt_fun} g(\mathbf{H}; \boldsymbol{\Theta}) &= \sum\limits_{n=1}^N\left\|\mathbf{h}_n - \mathbf{A}_\tau(\boldsymbol{\tau})\boldsymbol{\xi}_n\right\|^2.
\end{align}
The objective function \(g(\mathbf{H}; \boldsymbol{\Theta})\) is highly non-linear and no closed-form solution exists. Also, brute-force searching in this \(\mathbb{R}^{2LN+L}\) space is computationally intensive.
Therefore, the idea of EM is employed here to seek the solution iteratively.

Before the presentation of the EM approach for this problem, two issues need to be clarified:
\begin{enumerate}
\item As indicated by Fig. \ref{fig:calibration_flowchart}, at each CPP, the CFR can be sensed multiple times by consecutive UL-SRS symbols to improve the calibration accuracy. Here the symbol index, the number of total symbols, and each CFR measurement are denoted as \(q\), \(Q\), and \(\mathbf{H}^{(kq)}\), respectively. However, the following EM derivation drops the suffices of \(q\) and \(k\) for notational simplicity, which introduces no ambiguity since it is applied to each \(\mathbf{H}^{(kq)}\) individually.
  \item During the following derivation, we assume that the model order \(L\) is known. However, it is determined by the number of multipath components and is usually unavailable in real scenarios. That is why a path number estimation module is prepended to the estimation procedure as shown in Fig. \ref{fig:calibration_flowchart}, which can be implemented based on the information-theoretic criteria \cite{stoica2004_ModelorderSelection} or the sequential hypothesis-testing \cite{kritchman2009_NonParametricDetect}.
\end{enumerate}

Based on the notation of the EM algorithm \cite{feder1988_ParameterEstimation}, the observed CFRs \(\mathbf{h}_n, n = 1,\dots, N\) are named as the incomplete data since they are the amalgamation of \(L\) multipath components.
Then it is intuitive to choose the observations of each segregated path component
as the complete data, which are in the form of
\begin{equation}
  \mathbf{y}_{nl} = \mathbf{a}_\tau(\tau_l)\xi_{nl} + \mathbf{z}_{nl}, \quad l = 1,\dots,L,n = 1,\dots, N,
\end{equation}
where \(\mathbf{z}_{nl}\) denotes the noise components in the complete data, whose entries are i.i.d. with \(\mathcal{CN}\left(0, \beta_l\sigma_w^2\right)\), in which \(\sum_{l=1}^L\beta_l= 1\) and we choose \(\beta_l = \frac{1}{L}\) for simplicity.

The EM algorithm iteratively decomposes the observed incomplete signal \(\mathbf{h}_n\) into these segregated complete signals \(\mathbf{y}_{nl}\) and applies the MLE to obtain estimates of \(\boldsymbol{\Theta}\) from \(\mathbf{y}_{nl}\). These two steps are performed sequentially and iteratively based on the last estimates and are respectively referred to as the Expectation Step and Maximization Step. Denoting the estimates at the \(p\)-th iteration as \(\hat{\boldsymbol{\Theta}}^{(p)}\), then the \((p+1)\)-th iteration is carried out as follows.

\textbf{Expectation Step:}
\begin{align}
  \hat{\mathbf{y}}_{nl}^{(p+1)} &= \mathbf{a}_\tau(\hat{\tau}_l^{(p)})\hat{\xi}_{nl}^{(p)} + \frac{1}{L}\left[\mathbf{h}_n - \mathbf{A}_\tau(\hat{\boldsymbol{\tau}}^{(p)})\hat{\boldsymbol{\xi}}_n^{(p)}\right], \notag \\
  &\quad l = 1,\dots,L, n=1,\dots, N. \label{eq:expectation}
\end{align}

\textbf{Maximization Step:}
\begin{align}
\hat{\tau}_{l}^{(p+1)} &= \arg\max_{\tau_l}\left\{\sum_{n=1}^N\left|\mathbf{a}_\tau^{\mathsf{H}}(\tau_l)\hat{\mathbf{y}}_{nl}^{(p+1)}\right|^2\right\},\quad l=1,\dots,L, \label{eq:maximization_tau} \\
  \hat{\xi}_{nl}^{(p+1)} &= \frac{1}{M}\mathbf{a}_\tau^{\mathsf{H}}\left(\hat{\tau}_l^{(p+1)}\right)\hat{\mathbf{y}}_{nl}^{(p+1)},l=1,\dots,L,n=1,\dots,N.  \label{eq:maximization_xi}
\end{align}

As an iterative algorithm, the initial value of \(\boldsymbol{\Theta}\) and the stopping criteria have to be determined. Specifically, to reduce iteration numbers, the successive interference cancellation approach is adopted for initialization \cite{miridakis2013_SurveySuccessiveIn}.
Its main idea is to estimate the parameters of \(L\) paths successively and when estimating those of path \(l\), the interference caused by the previously estimated paths is calculated and subtracted from the CFR \(\mathbf{H}\). Each time when the interference caused by first \(l-1\) paths is canceled, the initial value for the parameters of path \(l\) is determined by correlating the interference-canceled CFR \(\mathbf{Y}\) with the delay signature function \(\mathbf{a}_\tau(\tau)\) and searching the peaks, as illustrated by the pseudo-code in Algorithm \ref{alg:SIC}.
To determine whether the EM iteration converges, the difference between consecutive estimates of \(\boldsymbol{\Theta}\) is calculated. When \(\left\|\hat{\boldsymbol{\tau}}^{(p)} - \hat{\boldsymbol{\tau}}^{(p - 1)}\right\| < \delta\tau\) and \(\frac{\left\|\hat{\boldsymbol{\xi}}_n^{(p)} - \hat{\boldsymbol{\xi}}_n^{(p - 1)}\right\|}{\left\|\hat{\boldsymbol{\xi}}_n^{(p - 1)}\right\|} < \epsilon, n = 1,\dots, N\) are met, the iteration stops, where \(\delta\tau\) is the search grid size for path delay and \(\epsilon\) is a pre-defined threshold.

\begin{algorithm}[htbp]
  \renewcommand{\algorithmicrequire}{\textbf{Input:}}
  \renewcommand{\algorithmicensure}{\textbf{Output:}}
  \caption{Successive interference cancellation for initialization of EM algorithm}
  \label{alg:SIC}
  \begin{algorithmic}[1]
    \REQUIRE CFR matrix \(\mathbf{H}\in\mathbb{C}^{M\times N}\) and path number \(L\).
    \STATE \(\mathbf{Y} \gets \mathbf{H}\);
    \FOR {\(l = 1 : L\)}
    \IF {\(l > 1\)}
    \STATE \(\mathbf{b} = [\hat{\xi}_{1(l-1)}, \dots, \hat{\xi}_{N(l-1)}]^{\mathsf{T}}\);
    \STATE \(\mathbf{Y} \gets \mathbf{Y} - \mathbf{a}_\tau(\hat{\tau}_{l-1}^{(0)})\cdot\mathbf{b}^{\mathsf{T}}\);
    \ENDIF
    \STATE \(\hat{\tau}_l^{(0)} = \argmax_\tau\left\|\left[\mathbf{a}_\tau^{\mathsf{H}}(\tau)\cdot\mathbf{Y}\right]^{\mathsf{T}}\right\|\);
    \STATE \(\left[\hat{\xi}_{1l}^{(0)}, \dots, \hat{\xi}_{Nl}^{(0)}\right] = \frac{1}{M}\mathbf{a}_\tau^{\mathsf{H}}(\hat{\tau}_l^{(0)})\cdot\mathbf{Y}\);
    \ENDFOR
    \ENSURE \(\hat{\boldsymbol{\Theta}}^{(0)}\).
  \end{algorithmic}
\end{algorithm}

As indicated by equations (\ref{eq:expectation}) to (\ref{eq:maximization_xi}), the EM algorithm simplifies the multi-dimensional maximum-likelihood search into iterative one-dimensional searches. Assuming \(J\) searching grids in the delay domain, then the computational complexities for the Expectation Step and the Maximization Step in each EM iteration are \(\mathcal{O}(MNL^2)\) and \(\mathcal{O}(JMNL) + \mathcal{O}(MNL)\), respectively. Since there are only \(6{\text- }8\) significant reflection paths in normal indoor environments \cite{czink2004_NumberMultipathClu, li2019_ClusterbasedChannel} and the RF channel numbers of small-cell base stations are usually no more than \(8\) (e.g. \(2\) or \(4\) for picocell RRUs) \cite{rodriguez2014_SmallCells5G}, \(L\) and \(N\) are small. Although wideband 5G signals occupy a large number of subcarriers, the CIR (inverse Fourier transform of the CFR) can be gated \cite{pan2022_EfficientJointDOAb} to reduce the effective subcarrier number based on the fact that 5G small-cell base stations have restricted power coverage. According to the analysis in our prior work (Section V-A of \cite{pan2022_EfficientJointDOAb}), for a coverage of \(100\;\mathrm{m}\), the subcarrier number \(M\) can be lowered to \(64\) after CIR gating. Therefore, it can be seen that \(N\approx L\ll M\ll J\) and the computational complexity of the EM iteration is dominated by the delay searching in the Maximization Step, which is \(\mathcal{O}(JMNL)\). Furthermore, the searching grid number \(J\) can also be substantially reduced by employing a coarse-to-fine searching strategy.

The EM iteration has been proven to be monotonically decreasing and has a fast convergence rate \cite{dempster1977_MaximumLikelihoodI}. Specifically, the EM solution for the MLE problem of (\ref{eq:NLMLE})-(\ref{eq:opt_fun}) generally converges within \(10\) iterations, according to our evaluations in typical multipath environments.  

Denoting the estimates of path delays and gains of the \(q\)-th CFR measurement at the \(k\)-th pilot position as \(\tilde{\tau}_{l}^{(kq)}\) and \(\tilde{\xi}_{nl}^{(kq)}\), respectively, then the shortest path with its gain larger than a threshold is picked as the LOS path. Here the paths with too small gains are filtered out by this threshold to guard against false local minima detected by the EM algorithm.

After that, as illustrated by Fig. \ref{fig:calibration_flowchart}, fixed phase rotations caused by path differences among antenna elements are subtracted from the path gain of the LOS path \(\tilde{\xi}_{n,\mathrm{LOS}}^{(kq)}\). This fixed phase rotation is determined by the array geometry and the known LOS DOA \(\theta_k\) of the calibration signal emitted from the \(k\)-th CPP, and after compensated, the path gain at the \(n\)-th antenna element is \(\xi_{n}^{(kq)} =\tilde{\xi}_{n, \mathrm{LOS}}^{(kq)}\cdot\left[\mathbf{a}_\theta'(\theta_k)\right]_n^*\). 

Next, the phase of \(\xi_{n}^{(kq)}\) is extracted as \(\phi_{n}^{(kq)} = \angle{\xi_{n}^{(kq)}}\), which represents the antenna phase error measurement at the \(n\)-th antenna for the direction of \(\theta_k\).
Taking into account that there exist sample timing offset, carrier frequency offset, and carrier phase offset in typical RF front-ends of commercial wireless communication equipment, the initial phase of UL-SRS varies across symbols \cite{mohammadian2021_RFImpairmentsWirel}.
Therefore, the differences between \(\phi_{n}^{(kq)}\) and \(\phi_1^{(kq)}\) are further calculated to derive a coherent measurement sequence over \(Q\) consecutive symbols. These phase differences are denoted as \(\Delta\phi_{n}^{(kq)}\) (obviously \(\Delta\phi_{1}^{(kq)} = 0\)) and they are combined to reduce the phase fluctuation caused by noise. For example, an outlier removal algorithm can be applied to this measurement sequence first, followed by retrieving the average value of the filtering results.

Up to this point, array phase error measurements at the discrete angular set \(\{\theta_k\}_{k = 1}^K\)  have been obtained (\(\Delta\bar{\phi}_{n}^{(k)}\)). Since DOA is a continuous variable, not necessarily an element in this angular set, the phase error function \(\varphi_n(\cdot)\) which can output the phase error value at any DOA needs to be inferred.
For this purpose, one can employ a parametric regression method, such as the polynomial curve fitting, to derive parameters of \(\varphi_n(\theta)\) directly, or a non-parametric regression method, such as the kernel regression or the interpolation, to obtain function values of \(\varphi_n(\theta)\) at pre-defined dense searching grids.

Lastly, based on the estimated phase error function \(\hat{\varphi}_n(\theta), n = 1,\dots, N\),
the array calibration process is performed by compensating the ideal array manifold \(\mathbf{a}_\theta'(\theta)\) with the estimated array modeling error function \(\hat{\boldsymbol{\zeta}}(\theta)\) as follows 
\begin{equation}
  \hat{\mathbf{a}}_\theta(\theta) = \mathbf{a}_\theta'(\theta) \odot \hat{\boldsymbol{\zeta}}(\theta).
\end{equation}
Since the calibrated array manifold \(\hat{\mathbf{a}}_\theta(\theta)\) captures different types of array errors and delineates the true array response for signals from any direction, DOA estimation using this matched array manifold achieves better performance than using the ideal but mismatched one (\(\mathbf{a}_\theta'(\theta)\)).

A wealth of searching-based DOA estimation algorithms, such as the conventional beamformer, the Capon method, the multiple signal classification method, the maximum likelihood estimators, and the compressive sensing-based methods, can directly employ this calibrated array manifold. They share a general DOA estimation procedure as follows:
\begin{enumerate}
\item Compute the calibrated array steering vectors on the predefined searching grids \(\{\theta_u\}_{u=1}^{U}\), which are actually the values of calibrated array manifold \(\hat{\mathbf{a}}_\theta(\theta)\) at the DOAs of \(\{\theta_u\}_{u=1}^{U}\).
\item Compute the spatial spectrum \(P(\theta_u), u = 1,\dots, U\) at these discrete grids using the calibrated steering vector set \(\{\hat{\mathbf{a}}_\theta(\theta_u)\}_{u=1}^U\).
\item Search the dominant spectral peaks and find the DOAs.
\end{enumerate}

Furthermore, in wireless positioning applications, to improve the degrees-of-freedom and resolution ability, JADE methods are usually employed to estimate the DOA and TOA simultaneously rather than separately \cite{vanderveen1997_JointAngleDelay}. Similar to the idea of DOA estimation presented above, they can also use the calibrated array manifold for spatial processing to counteract array errors. One can refer to our prior works \cite{pan2022_EfficientJointDOAb} and \cite{pan2021_JointDoAToA} for detailed discussions about how to use the calibrated array manifold in different JADE methods.

\section{Numerical Simulations}
\label{sec:simulation}
\subsection{Simulation Setup}
In this section, the effectiveness of the proposed in-situ calibration framework is demonstrated with simulated 5G wireless channel data.
The system and waveform parameters used in simulations are shown in TABLE \ref{tab:exp_parameters}.

\begin{table}[htbp]
  \caption{Configurations for 5G system and UL-SRS}
  \begin{center}
    \begin{tabular}{cc}
      \toprule
      \bf{Parameter} & \bf{Value} \\
      \midrule
      Type of gNB & Picocell gNB \\
      Number of antenna elements in an RRU & \(4\) \\
      Array type & ULA \\
      Carrier frequency & \(4.85\;\text{GHz}\) \\
      Subcarrier spacing & \(30\;\text{kHz}\) \\
      Number of subcarriers & \(3264\) \\
      UL-SRS pattern & Comb-two \cite{3gpp.38.211}  \\
      UL-SRS transmission bandwidth & \(100\;\text{MHz}\) \\
      Sampling frequency & \(122.88\;\text{MHz}\) \\
      UL-SRS temporal interval & \(80\;\text{ms}\) \\
      \bottomrule
    \end{tabular}
    \label{tab:exp_parameters}
  \end{center}
\end{table}
\vspace{-1.6em}

\begin{figure}[htbp]
  \centering
  \includegraphics[width=0.35\textwidth]{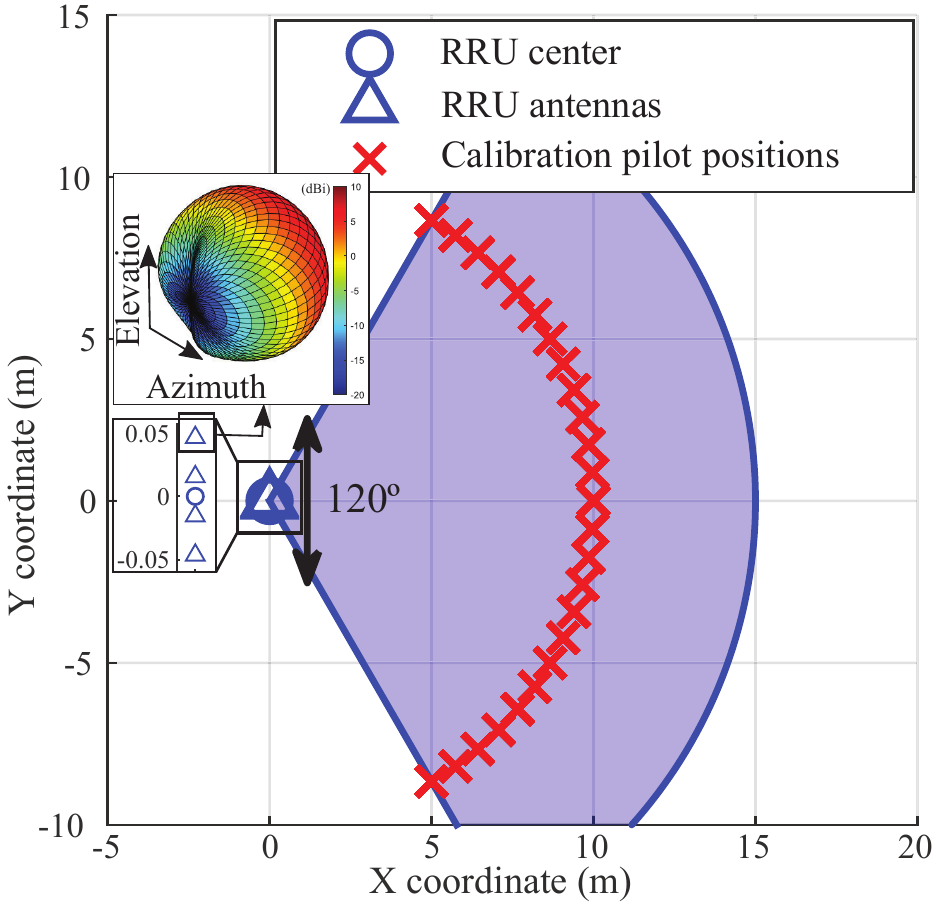}
  \caption{Visualization of simulated scenario for in-situ calibration.}
  \label{fig:calibration_pilots_plot}
\end{figure}

To test in-situ antenna array calibration methods in typical multipath environments,
simulations are conducted with realistic wireless channel data generated by the QuaDRiGa channel simulator \cite{jaeckel2014_QuaDRiGa3DMultiCel}.
The indoor factory LOS channel at sub-6GHz working frequency (\(\mathsf{3GPP\_38.901\_InF\_LOS}\)) whose parameters conform to \cite{3gpp.38.901} is chosen for QuaDRiGa throughout the experiments.
The antenna element pattern is configured according to the default antenna modeling parameters defined in this same 3GPP report \cite{3gpp.38.901}.
Specifically, its \(3\;\mathrm{dB}\) beamwidths in both azimuth and elevation directions are set to \(65^\circ\) and the directional antenna gain is set to \(8\;\mathrm{dBi}\), as illustrated by its 3-D radiation pattern in Fig. \ref{fig:calibration_pilots_plot}.
To simulate the direction-dependent array errors, we modulate each multipath component generated by the QuaDRiGa simulator with an additional phase offset, whose value is determined by its DOA and a look-up-table with antenna phase measurements of a realistic four-element antenna array equipped by a 5G RRU.   
During simulations, the transmit power of the UT and the noise figure of the gNB are fixed to \(P_t = 200\;\mathrm{mW}\) and \(F = 5\;\mathrm{dB}\), respectively.
Then the power of the receiving signal \(P_r\) is derived by the QuaDRiGa simulator according to the propagation model and the noise power is calculated as \(P_n = k_BT_0B\), in which  \(k_{\mathrm{B}}\), \(T_0\), and \(B\) represent the Boltzmann's constant, standard noise temperature, and measurement bandwidth, respectively. The noise component on each subcarrier of each receiving channel is generated independently according to the complex Gaussian distribution of \(\mathcal{CN}(0, P_n)\).

The performance of the proposed EM-based array manifold estimation method is compared against: (i) the widely adopted PE-based approach \cite{see1995_MethodArrayCalibra, lanne2006_CalibratingArraySc, heidenreich2009_HighresolutionDirec, ibanezurzaiz2021_DigitalBeamforming}, and (ii) the direct measuring approach \cite{vasanelli2020_CalibrationDirectio, pan2022_EfficientJointDOAb}. The latter approximates the array phase response by the measured phases of the multi-channel CFR at the center frequency. Since a clean one-path wireless environment is assumed for this method, it is commonly used in chamber calibration.
These three array manifold estimation methods are all applied in the proposed in-situ calibration framework as presented in Section \ref{sec:insitu_framework}.
Their performance comparisons are presented in this section according to the metrics of (i) the accuracy of the estimated array manifold and (ii) the DOA estimation error of the calibrated array.

Moreover, to evaluate the in-situ calibration performance
in different multipath environments, the Ricean K-factors for the simulated wireless channels are configured to vary from \(0\;\mathrm{dB}\) to \(7\;\mathrm{dB}\) during simulations.

\subsection{Evaluating Accuracy of Estimated Array Manifold}
\label{sec:eval_manifold}
We first evaluate the accuracy of the estimated array manifold by comparing it to the true manifold. We place the UT at \(25\) evenly distributed pilot positions on an arc centered at the RRU, as indicated by cross marks in Fig. \ref{fig:calibration_pilots_plot}. 
Their distances to the RRU are \(10\;\mathrm{m}\) and they cover the sector of \(120\) degrees (\(-60^\circ\) to \(+60^\circ\)) with an angular separation of \(5\) degrees.
That is \(\theta_k = -60 + 5(k-1), k = 1,\dots,K\), where the total CPP number \(K\) is \(25\). Then with the Ricean K-factor and CPP fixed, \(500\) UL-SRS symbols are simulated by the QuaDRiGa channel simulator.

Channel responses sensed by the RRU for these \(25\) pilot positions are used for array manifold estimation.
The angle between the estimated manifold \(\hat{\mathbf{a}}_\theta(\theta)\) and the true manifold \(\mathbf{a}_\theta(\theta)\) at a specific direction \(\theta\), which is essentially the angle between two vectors, represents the manifold mismatch and is used as the metric for evaluating the accuracy of the estimated array manifold. It is calculated as 
\begin{equation}
  \alpha(\theta) = \mathrm{arccos}\left[\frac{\left|\hat{\mathbf{a}}_\theta^{\mathsf{H}}(\theta)\mathbf{a}_\theta(\theta)\right|}{\left\|\hat{\mathbf{a}}_\theta(\theta)\right\|\cdot\left\|\mathbf{a}_\theta(\theta)\right\|}\right].
\end{equation}

\begin{figure*}[htbp]
  \centering
  \subfloat[Manifold mismatches in different multipath conditions when a single UL-SRS symbol is used.]{\centering\includegraphics{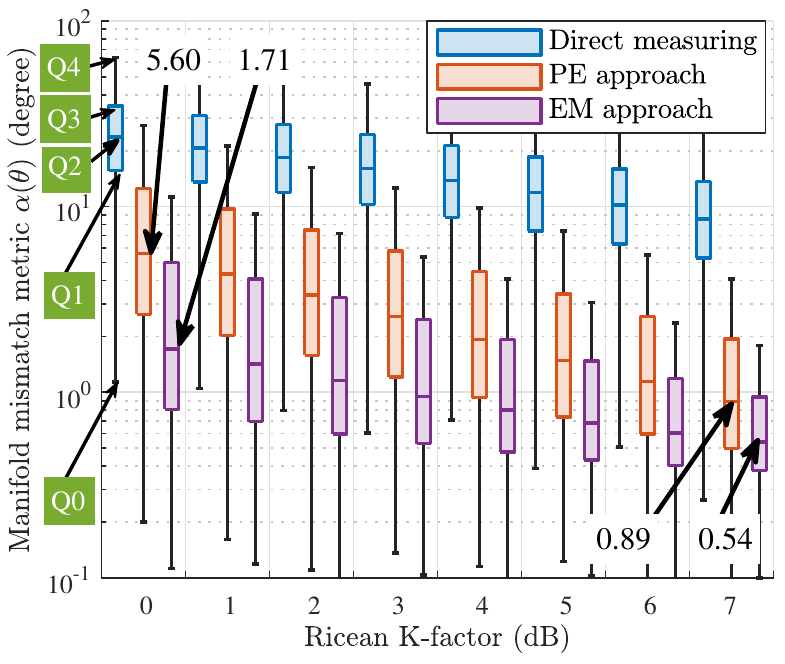}}
  \quad\quad\quad
  \subfloat[Manifold mismatches when multiple UL-SRS symbols are used (Ricean K-factor fixed to \(3\;\mathrm{dB}\)).]{\centering\includegraphics{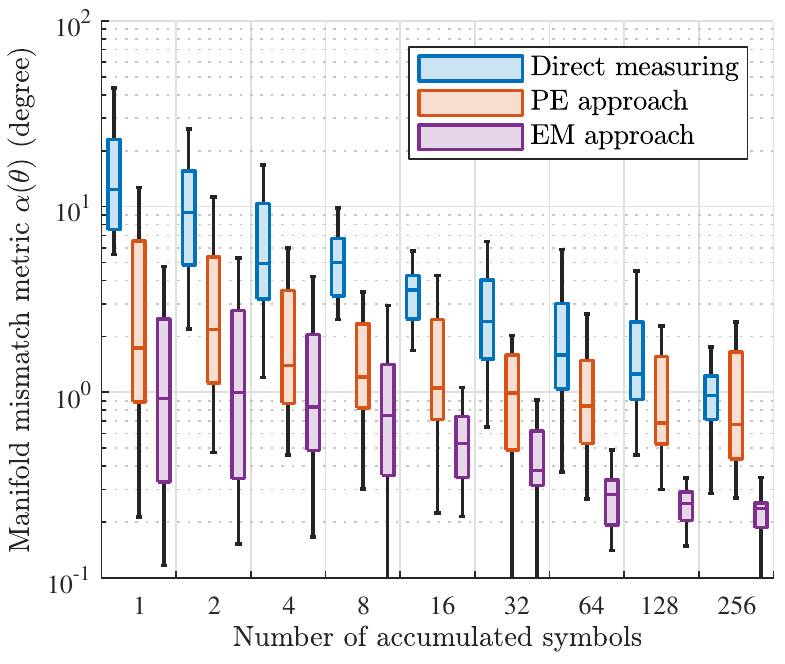}}
  \caption{Demonstration of manifold mismatches of the proposed EM-based array manifold estimation algorithm and benchmark algorithms under different multipath conditions and different numbers of accumulated UL-SRS symbols.}
  \label{fig:manifold_mismatch_compare}
\end{figure*}

Two sets of experiments are conducted to investigate the impacts of the multipath condition and the number of accumulated UL-SRS symbols on the array manifold estimation performance, respectively. Their results are shown in Fig. \ref{fig:manifold_mismatch_compare}.

First, to assess the estimation performance under the extreme single
snapshot scenario, the array manifold is estimated based on a single
CFR measurement. We use the standard box plot (a.k.a.
  the box-and-whisker plot) to visualize the statistics of
  \(\alpha(\theta)\) as it is more informative when used for error
  analysis than single-metric evaluations such as the \(N\)-th
  percentile of the error set or the root mean square error
  \cite{boxplots}. The resulting box plots of
\(\alpha(\theta_k), k=1,\dots,K\) at each Ricean K-factor are shown in
Fig. \ref{fig:manifold_mismatch_compare}(a) for the proposed and
benchmark methods. As stated above, each box plot
demonstrates the statistics of \(\alpha(\theta)\) from
\(25\times500=12500\) realizations. As exemplified by
  Fig. \ref{fig:manifold_mismatch_compare}(a), the minimum (\(Q_0\) or
  \(0\)-th percentile), first quartile (\(Q_1\) or \(25\)-th
  percentile), median (\(Q_2\) or \(50\)-th percentile), third
  quartile (\(Q_3\) or \(75\)-th percentile), and maximum (\(Q_4\) or
  \(100\)-th percentile) of the \(\alpha(\theta)\) set are illustrated
  in the corresponding box plot, as respectively represented by the
  lower limit of the lower whisker, the lower edge of the box, the
  middle line of the box, the upper edge of the box, and the upper
  limit of the upper whisker.

Fig. \ref{fig:manifold_mismatch_compare}(a) shows that the direct measuring approach performs poorly in the presence of multipaths. It also clearly indicates the superiority of the proposed EM-based manifold estimation method over the PE-based approach, especially in dense multipath environments. For example, when the Ricean K-factor is \(0\;\mathrm{dB}\),
which denotes a severe multipath condition with the averaged signal power from scattered paths equals to the LOS signal power,
a reduction of \(69\%\) of the median of \(\alpha(\theta)\) is achieved by the EM-based approach (from \(5.60^\circ\) to \(1.71^\circ\)); While when the Ricean K-factor is \(7\;\mathrm{dB}\), the performance improvement is only \(39\%\) (median error reduces from \(0.89^\circ\) to \(0.54^\circ\)).
This implies that, limited by the spatial resolution of the small-scale antenna array, conventional spatial-domain-only in-situ calibration methods
exhibit unsatisfactory performance in the presence of multipath reflections.
In contrast, the proposed approach resolves multipaths via delay-domain super-resolution, and the large bandwidth of 5G signals guarantees an accurate estimation of the real antenna manifold even in a multipath-rich environment. 

Then we examine the performance improvement for these array manifold estimation algorithms when multiple UL-SRS symbols are used.
To fully utilize these multiple measurements, the outlier rejection algorithm based on the Hampel identifier \cite{pearson2016_GeneralizedHampelF} is applied to the corresponding phase estimates and the filtering results are averaged to derive the final array manifold estimate at this CPP.
In this experiment, the Ricean K-factor of the wireless channel is fixed to \(3\;\mathrm{dB}\) and the number of accumulated symbols varies from \(1\) to \(256\). Fig. \ref{fig:manifold_mismatch_compare}(b) shows the box plots of manifold mismatches for the three array manifold estimation methods, with each box demonstrating the statistic of \(\alpha(\theta_k)\) at all \(25\) CPPs. 

We readily observe from Fig. \ref{fig:manifold_mismatch_compare}(b) that all methods benefit from multiple measurements and the EM-based approach is superior to both the benchmark algorithms in all these multi-snapshot scenarios.
It also shows that, even when there are only \(4\) symbols, the median estimation error is below \(0.9^\circ\) for the proposed EM-based approach. This implies that the time and effort required for measuring during in-situ calibration can be saved by reducing the dwell time at each CPP, or more conveniently, a UT travels across the angular coverage of the RRU can be used to provide continuous measurements during which the samplings of the UT trajectory at the UL-SRS transmitting instants form the CPPs. 

\begin{figure*}[htbp]
  \centering
  \subfloat[Ricean K-factor: \(0\;\mathrm{dB}\) (PE approach).]{\includegraphics[width=0.33\textwidth]{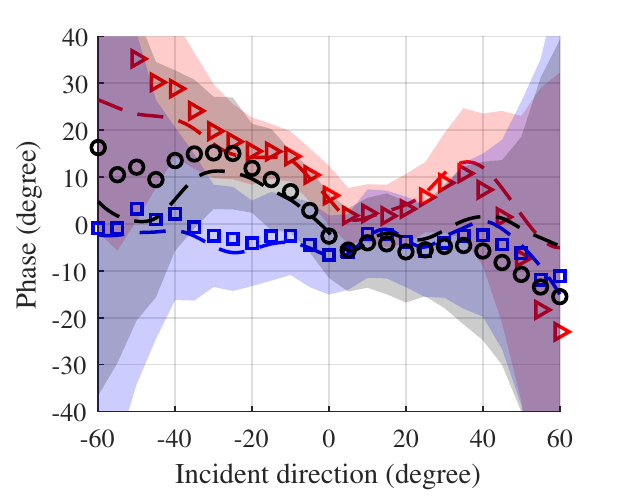}}
  \hfill
  \subfloat[Ricean K-factor: \(3\;\mathrm{dB}\) (PE approach).]{\includegraphics[width=0.33\textwidth]{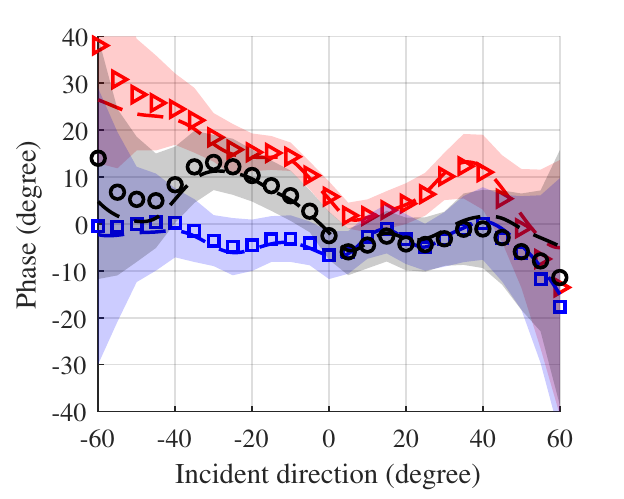}}
  \hfill
  \subfloat[Ricean K-factor: \(7\;\mathrm{dB}\) (PE approach).]{\includegraphics[width=0.33\textwidth]{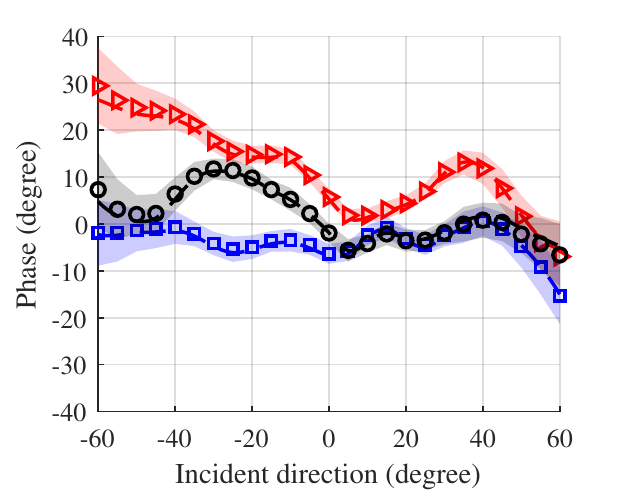}}\\
  \subfloat[Ricean K-factor: \(0\;\mathrm{dB}\) (EM approach).]{\includegraphics[width=0.33\textwidth]{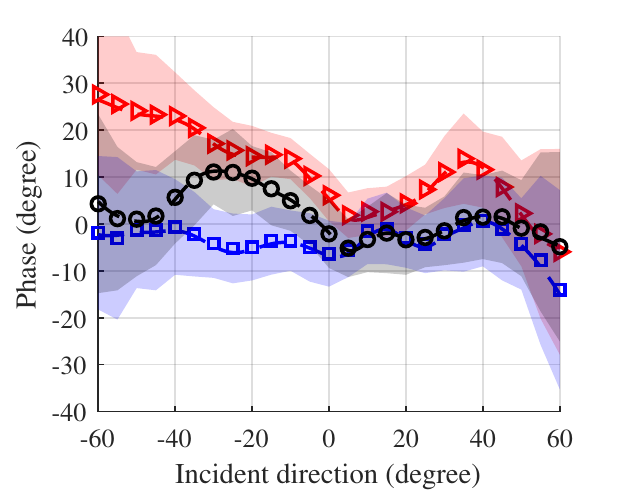}}
  \hfill
  \subfloat[Ricean K-factor: \(3\;\mathrm{dB}\) (EM approach).]{\includegraphics[width=0.33\textwidth]{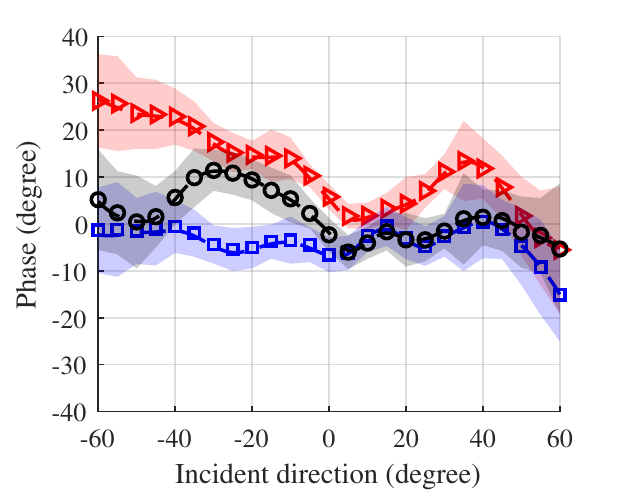}}
  \hfill
  \subfloat[Ricean K-factor: \(7\;\mathrm{dB}\) (EM approach).]{\includegraphics[width=0.33\textwidth]{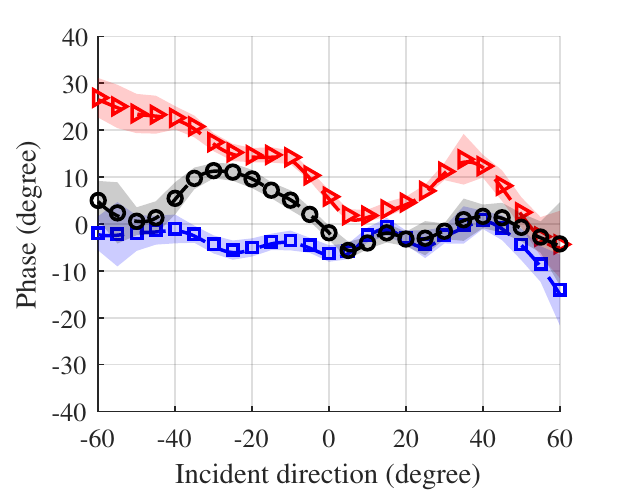}}\\
  \subfloat{\includegraphics{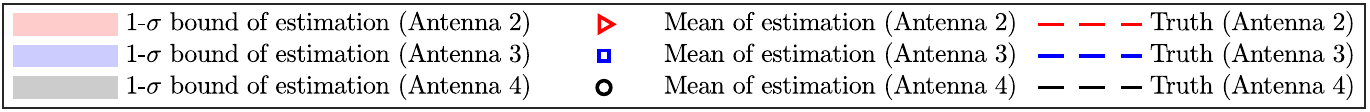}}
  \caption{Demonstration of antenna phase errors at the directions of CPPs estimated by the PE-based approach ((a)-(c)) and the proposed EM-based approach ((d)-(f)).}
  \label{fig:phase_estimation}
\end{figure*}

Further, Fig. \ref{fig:phase_estimation} demystifies the underlying antenna phase error estimates when the Ricean K-factors are \(0\;\mathrm{dB}\), \(3\;\mathrm{dB}\), and \(7\;\mathrm{dB}\), respectively. Here, we only show the estimates obtained by the PE-based and the proposed EM-based approaches as the direct measuring method exhibits much higher estimation variance, which is obvious from Fig. \ref{fig:manifold_mismatch_compare}.
The mean and variance of array phase error estimates from \(500\) CFR measurements at each CPP are shown in Fig. \ref{fig:phase_estimation}.
It can be observed from Fig. \ref{fig:phase_estimation} that the proposed EM-based array manifold estimator outperforms that based on the PE in terms of stability and accuracy. The increased \(1{\text -}\sigma\) bounds for both methods in large incident directions, as shown in Fig. \ref{fig:phase_estimation}, are attributed to the decreased receiving power of the directive antenna elements for signals impinged from these directions.  

\subsection{Evaluating DOA Estimation Error of Calibrated Array}
\label{sec:simu_eval_doa}
We then demonstrate the performance improvements for DOA estimation of different in-situ calibration techniques.
To this end, DOA estimation is performed on the simulated 5G channel data with the estimated array manifolds.
We conduct \(1000\) Monte Carlo trials under each multipath condition and in each trial, the path DOA, TOA, and the wireless channel response are generated randomly.
Specifically, the DOA and TOA of the LOS path conform to \(\mathcal{U}\left(-60^\circ, +60^\circ\right]\) and \(\mathcal{U}\left(0, 333.33\;\mathrm{ns}\right]\), respectively, which means the UT locates in a sector with a central angle of \(120^\circ\) and a radius of \(100\;\mathrm{m}\) centered at the RRU. 

While Section \ref{sec:eval_manifold} only investigates the array manifold estimates at the discrete CPPs, a continuous or a finer array manifold is needed for DOA estimation. Therefore, following the flowchart of Fig. \ref{fig:calibration_flowchart}, we smooth the discrete phase error estimates with the local weighted regression \cite{clevelandRobustLocallyWeighted1979} and interpolate the regression results with the Akima spline \cite{akima1970_NewMethodInterpola} to derive a continuous phase error curve, which is then utilized to obtain the continuous manifold. Besides, to handle the multipath effects during DOA estimation, we employ the JADE method proposed in our prior work \cite{pan2022_EfficientJointDOAb} for parameter estimation.
It composes of an iterative-adaptive-approach-based delay spectrum estimator and a conventional beamformer. In addition, the DOA estimates with non-calibrated and perfectly calibrated manifolds are also investigated for comparison.

\begin{figure*}[htb]
  \centering
  \subfloat[\(80\)-th percentiles of DOA estimation errors under different multipath conditions (Number of UL-SRS symbols used in array manifold estimation is \(8\)).]{\includegraphics{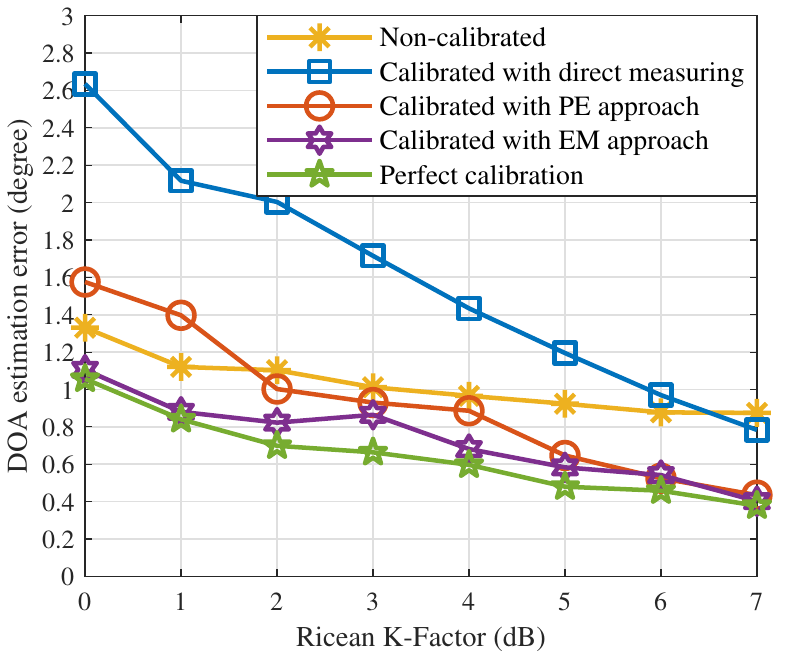}}
  \quad\quad\quad
  \subfloat[\(80\)-th percentiles of DOA estimation errors using manifolds estimated with different numbers of UL-SRS symbols (Ricean K-factor fixed to \(3\;\mathrm{dB}\)).]{\centering\includegraphics{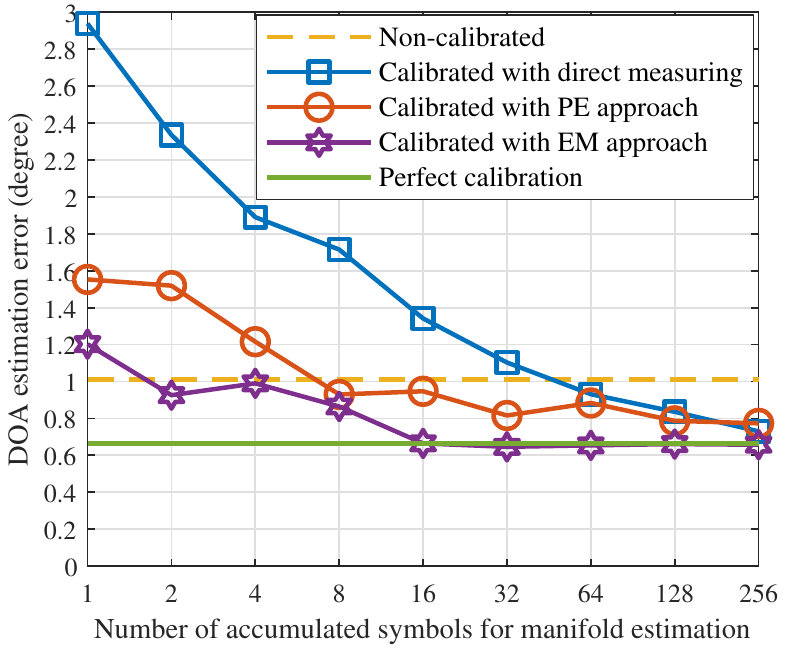}}
  \caption{DOA estimation errors when using array manifolds estimated by the proposed EM-based algorithm and benchmark algorithms under different multipath conditions and when different numbers of UL-SRS symbols are accumulated for manifold estimation.}
  \label{quadriga_doa}
\end{figure*}

Similar to the methodology adopted in Section \ref{sec:eval_manifold}, the DOA estimation errors with different estimated array manifolds are compared by varying the multipath condition and the number of accumulated UL-SRS symbols.
The corresponding results are summarized in Fig. \ref{quadriga_doa}(a) and Fig. \ref{quadriga_doa}(b), respectively.

First, Fig. \ref{quadriga_doa}(a) shows the \(80\)-th percentiles of DOA estimation errors when the Ricean K-factor varies from \(0\;\mathrm{dB}\) to \(7\;\mathrm{dB}\). The count of UL-SRS symbols accumulated for array manifold estimation is fixed to \(8\) throughout this experiment. To simulate the scenario that in-situ calibration and the subsequent positioning process are carried out in the same environment, the array manifolds used for DOA estimation are estimated under the same Ricean K-factor.

Fig. \ref{quadriga_doa}(a) illustrates that, in terms of DOA estimation error, the performance of the proposed EM-based in-situ calibration method approximates that of the perfect calibration when the Ricean K-factor is no less than \(0\;\mathrm{dB}\), while a similar effect is achieved by the traditional PE-based method only when the Ricean K-factor is above \(5\;\mathrm{dB}\). Also according to the results, calibrated with the directly in-situ measured manifold facilitates DOA estimation only when the measurement process is conducted in a clear wireless environment (when the Ricean K-factor is above \(6\;\mathrm{dB}\) as shown in Fig. \ref{quadriga_doa}(a)).

Next in Fig. \ref{quadriga_doa}(b), we fix the Ricean K-factor to \(3\;\mathrm{dB}\) and demonstrate the DOA estimation errors when using array manifolds estimated with different numbers of UL-SRS symbols. Results shown in Fig. \ref{quadriga_doa}(b) confirm the superiority of the proposed EM-based method over both the benchmark algorithms when multiple snapshots exist. We can also observe that it achieves nearly identical performance as the perfect calibration when at least \(16\) CFR measurements are utilized for manifold estimation. By contrast, although the estimation errors of both the benchmark in-situ calibration methods decrease as the snapshot number increases, they can hardly approach that of perfect calibration. 

\section{Indoor Field Tests}
\label{sec:field_tests}
\subsection{Experimental Setup}
\begin{figure*}[htbp]
  \centering
  \includegraphics[width=0.8\textwidth]{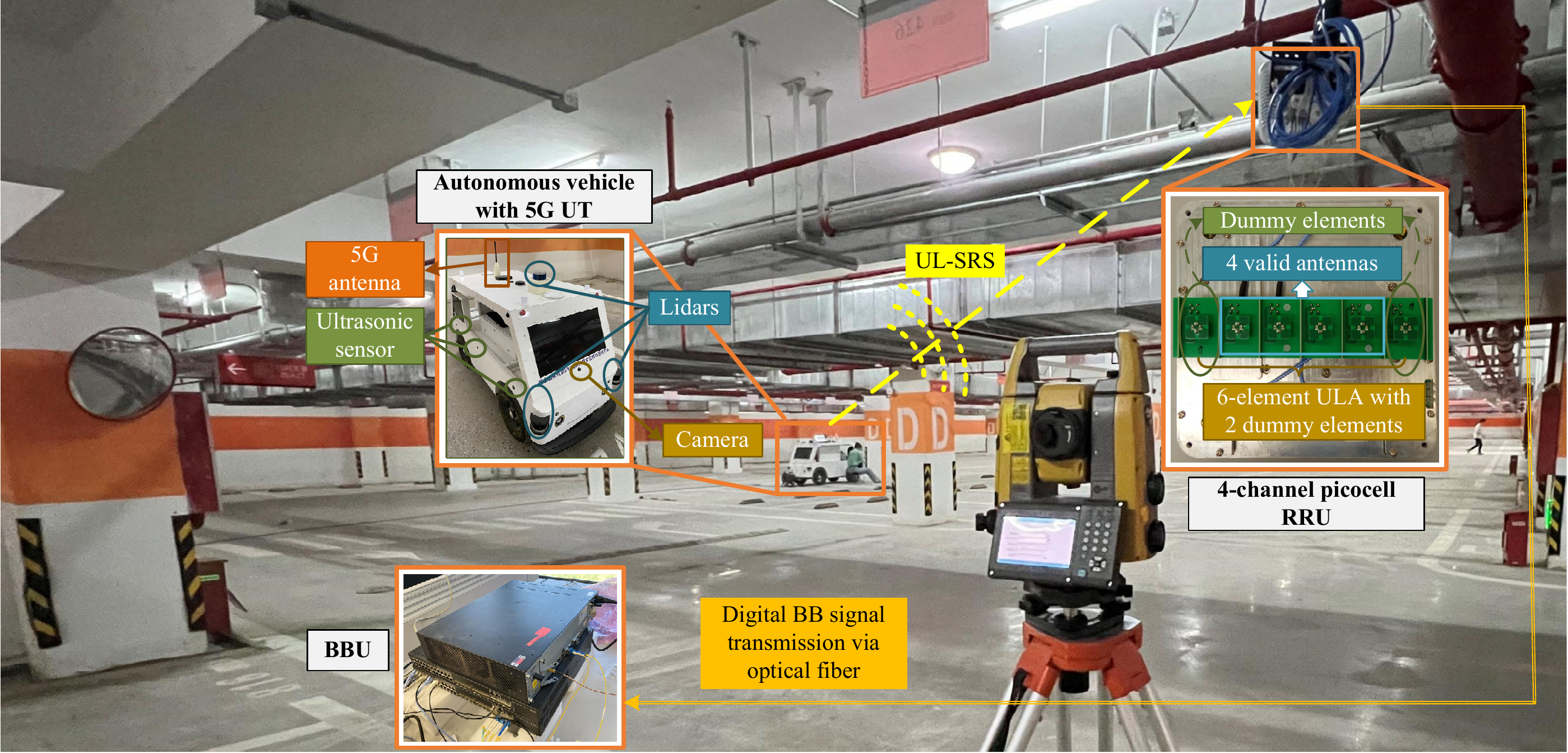}
  \caption{Experimental environment and hardware setups for field test in an underground parking lot.}
  \label{fig:hardware_setup}
\end{figure*}

To further verify the effectiveness of the proposed in-situ calibration method and demonstrate its performance improvement for DOA estimation, field tests are conducted in an underground parking lot of an office building\footnote{The field test data used in performance evaluations is available in \cite{k2f0-k132-22}.}.  
Fig. \ref{fig:hardware_setup} depicts the experimental environment and hardware setups. During experiments, we only use a single RRU and a single UT, whose working parameters also conform to TABLE \ref{tab:exp_parameters}. As shown in Fig. \ref{fig:hardware_setup}, this area has a lot of metallic plumbing pipes, poles, and thick pillars, which cause harsh multipath effects for wireless signals.

\begin{figure*}[htb]
  \centering
  \subfloat[Layout and photograph of the designed antenna array.]{\includegraphics[width=0.36\textwidth]{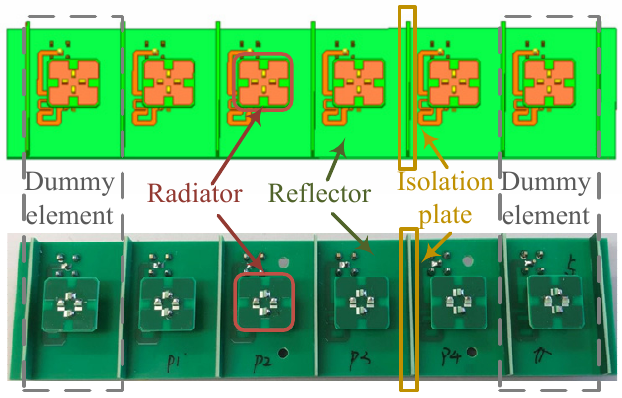}}\hfill
  \subfloat[3-D radiation pattern of the antenna element.]{\includegraphics[width=0.32\textwidth]{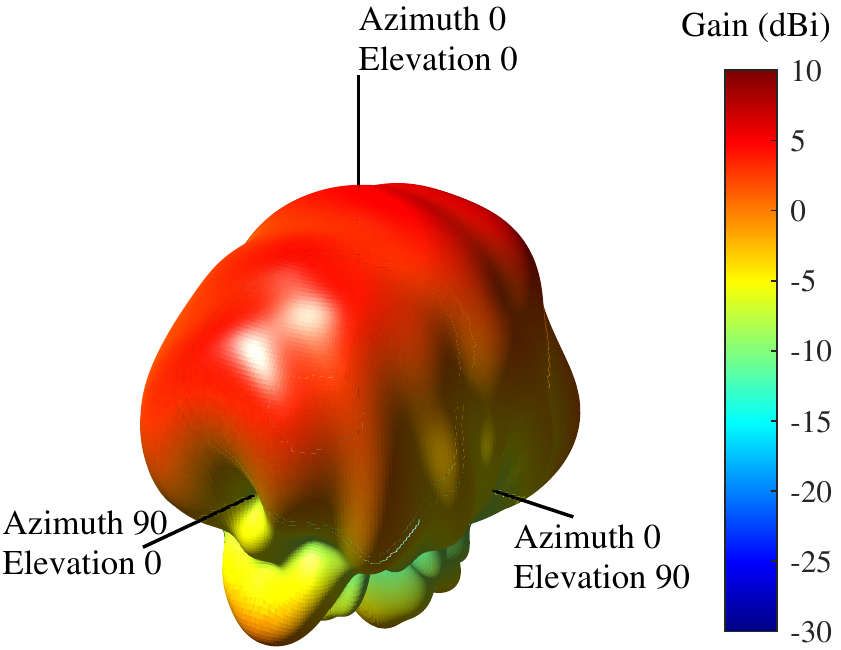}}\hfill
  \subfloat[2-D radiation patterns of the antenna element in H-plane and E-plane.]{\includegraphics[width=0.255\textwidth]{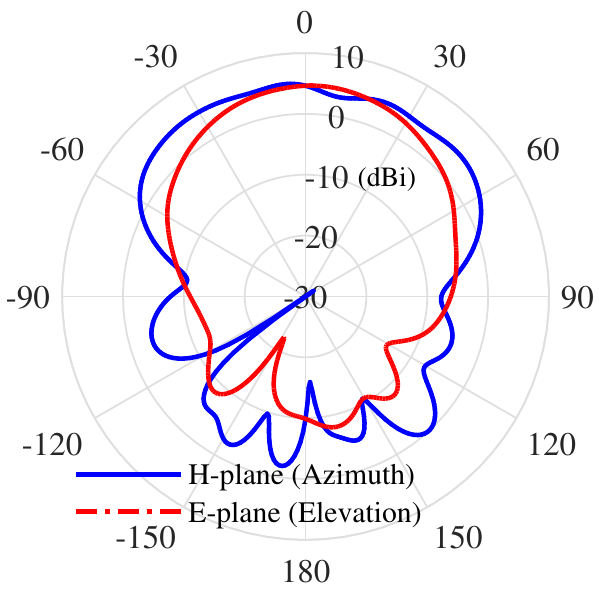}}
  \caption{Layout, photograph, and radiation patterns of the antenna array.}
  \label{fig:antenna_array}
\end{figure*}

Main devices used in experiments have been illustrated in Fig. \ref{fig:hardware_setup}, and they are summarized as follows:  
\begin{enumerate}
\item UT: The 5G UT with a single omnidirectional cylindrical antenna is mounted on an autonomous vehicle, which is also equipped with various active sensors, including an inertia measurement unit and multiple lidars, cameras, and ultrasonic distance sensors. Measurements from these active sensors are fused via a simultaneous localization and mapping algorithm to generate ground-truth locations with an accuracy of several centimeters.
\item RRU: We use a commercial four-channel picocell RRU in field tests, which incorporates the RF and IF processing modules shown in Fig. \ref{fig:measurement_setup}. Specifically, RF signals are first filtered, amplified, down-converted, and sampled to digital IF signals. Then digital down converters are followed to generate BB in-phase and quadrature signals. The original antennas equipped by this RRU are dispersed at four corners with element spacing much larger than half-wavelength, preventing it from supporting the DOA estimation function. Therefore, we designed and fabricated a six-element ULA to replace the existing RRU antennas. Its middle four antennas connect to the RRU RF channels accordingly, while those at both sides are dummy elements, which guarantee the same boundaries seen by the central four elements of the array \cite{raeesi2022_BidirectionalMEMSLi}. The element spacing is \(3\;\mathrm{cm}\). Fig. \ref{fig:antenna_array}(a) presents the structure of this antenna array. The simulated 3-D and 2-D radiation patterns of the antenna element are shown in Fig. \ref{fig:antenna_array}(b) and Fig. \ref{fig:antenna_array}(c), respectively. The corresponding specifications are also listed in TABLE \ref{tab:ant_parameters}. 
\item BBU: A commercial 5G BBU is employed, whose physical layer modules are fully compatible with the 3GPP standard. As shown in Fig. \ref{fig:measurement_setup}, during in-situ calibration, it processes BB UL-SRS signals and outputs CFR measurements. It also needs to mention that the BBU is placed in the equipment room rather than in the experimental field and is connected to the RRU via a long optical fiber. 
\end{enumerate}

\begin{table}[htbp]
  \caption{Specifications of antenna elements of the designed antenna array}
  \begin{center}
    \begin{tabular}{cc}
      \toprule
      \bf{Parameter} & \bf{Value} \\
      \midrule
      Type & Microstrip antenna \\
      Working frequency range & \(4.80{\text -}4.90\;\mathrm{GHz}\) \\
      Polarization & Vertical polarization \\
      Gain & \(5.20\;\mathrm{dBi}\) at \(4.85\;\mathrm{GHz}\) \\
      Efficiency & \(93\%\) at \(4.85\;\mathrm{GHz}\) \\
      H-plane HPBW & \(122^\circ\) at \(4.85\;\mathrm{GHz}\) \\
      E-plane HPBW & \(72^\circ\) at \(4.85\;\mathrm{GHz}\)\\
      VSWR & less than \(1.5\) in \(4.80{\text -}4.90\;\mathrm{GHz}\) \\
      \bottomrule
    \end{tabular}
    \label{tab:ant_parameters}
  \end{center}
\end{table}

\subsection{Experiment Results}
\begin{figure}[htb]
  \centering
  \includegraphics{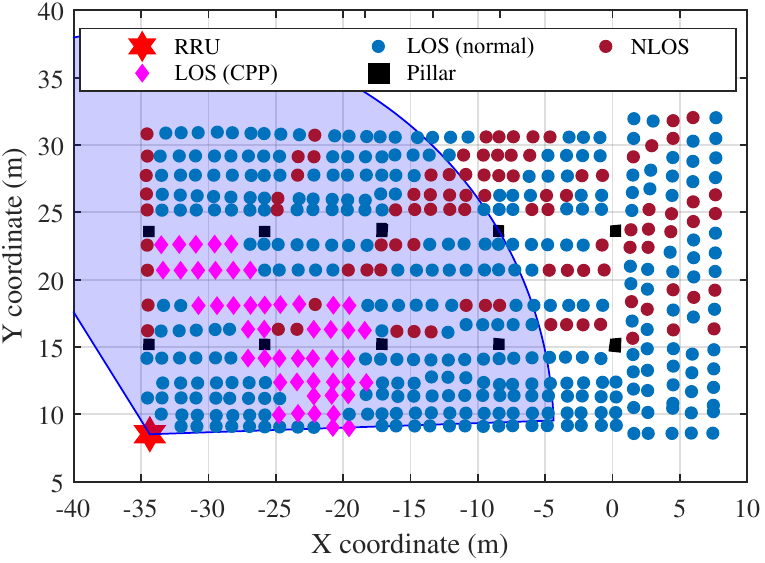}
  \caption{Experiment layout and visualizations for data-collecting positions.}
  \label{fig:loc_visual}
\end{figure}

In the field test,
the RRU is fixed in position \((-34.4\;\mathrm{m}, 8.5\;\mathrm{m})\) and
the UT moves in a rectangular area of nearly \(1125\;\mathrm{m}^2 \left([-35\;\mathrm{m}, 10\;\mathrm{m}] \times [8\;\mathrm{m}, 33\;\mathrm{m}]\right)\) and stops at \(476\) coordinates in this area with an interval of about \(1.5{\text -}2.5\;\mathrm{m}\), as depicted in Fig. \ref{fig:loc_visual}.
Among them, \(48\) LOS positions around the RRU are chosen as the CPPs and channel measurements at those positions are used for array manifold estimation. Also as shown in Fig. \ref{fig:loc_visual}, ten evenly spaced pillars exist in this rectangular area, which give rise to \(95\) NLOS positions. Therefore, CFRs collected at the remaining \(333\) normal LOS positions are used for DOA estimation performance evaluations. At each position, the UT sends \(100\) UL-SRS symbols.

\begin{figure}[htb]
  \centering
  \subfloat[Phase error at antenna \(2\).]{\includegraphics{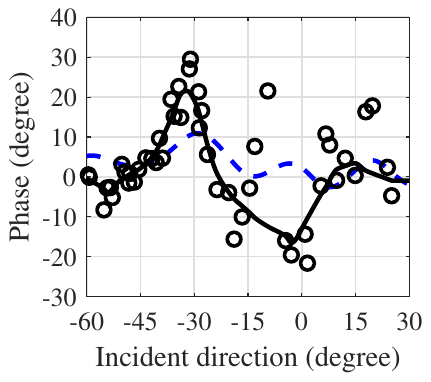}}
  \hfill
  \subfloat[Phase error at antenna \(3\).]{\includegraphics{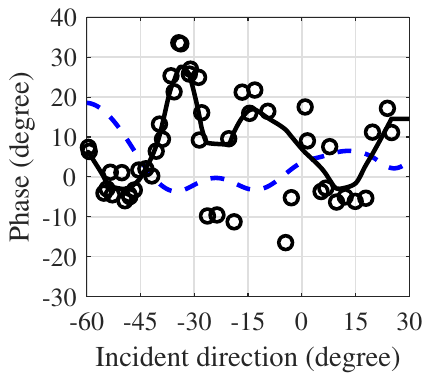}}\\
  \subfloat[Phase error at antenna \(4\).]{\includegraphics{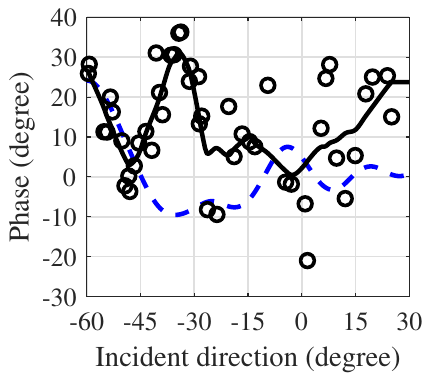}}\hfill\subfloat{\includegraphics{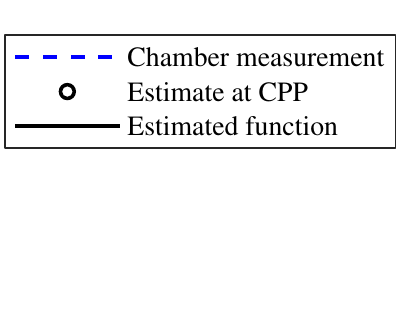}}
  \caption{Demonstration of array phase errors estimated by the proposed EM-based approach at CPPs and the estimated phase error functions \(\varphi_n(\theta), n = 2,3,4\). Since \(\varphi_1(\theta) \equiv 0\), (a), (b), and (c) respectively show the corresponding estimates for the antenna elements \(2\), \(3\), and \(4\).}
  \label{fig:fieldtest_ant_phs_err}
\end{figure}

Fig. \ref{fig:fieldtest_ant_phs_err} demonstrates the phase error estimates obtained by the proposed EM-based method at these \(48\) CPPs.
Only \(10\) UL-SRS symbols are used at each CPP for phase error estimation. To derive the continuous phase error function for a specific antenna element, the same approach stated in Section \ref{sec:simu_eval_doa} is also used here, which encompasses the local weighted regression \cite{clevelandRobustLocallyWeighted1979} and the Akima spline interpolation \cite{akima1970_NewMethodInterpola}.
The deviation of the in-field array manifold from the nominal manifold measured in an anechoic chamber is clearly illustrated. 

\begin{figure}[htb]
  \centering
  \subfloat[Empirical CDF curves for DOA estimation errors.]{\includegraphics{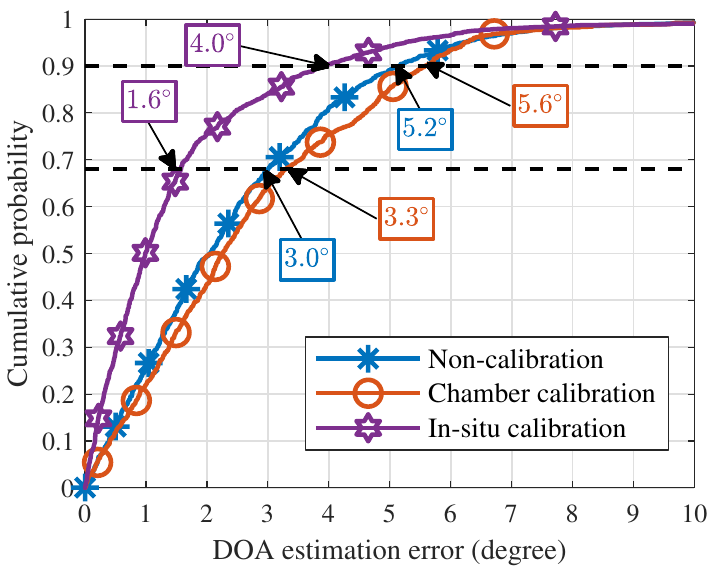}}\\
  \subfloat[Box plots of DOA estimation errors in different sections of incident directions.]{\includegraphics{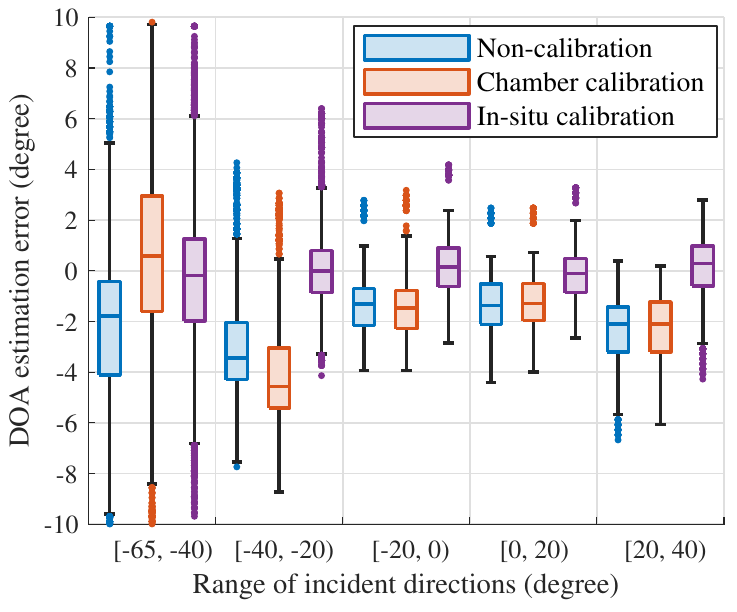}}
  \caption{Comparisons of DOA estimation performance with non-calibrated, chamber calibrated, and in-situ calibrated array manifolds.}
  \label{fig:doa_performance}
\end{figure}

The ideal array manifold is calibrated by the chamber-measured and in-situ estimated phase error functions shown in Fig. \ref{fig:fieldtest_ant_phs_err} to obtain the nominal and estimated array manifolds.
Then the ideal, nominal, and estimated array manifolds are used to estimate the DOAs of signals transmitted from the \(333\) normal LOS positions (The total number of samples is \(333\times100=33300\)).
Their corresponding results are respectively denoted as non-calibration, chamber calibration, and in-situ calibration results.
Similar to simulations presented in Section \ref{sec:simu_eval_doa}, the JADE method proposed in \cite{pan2022_EfficientJointDOAb} is also adopted here for DOA estimation. 
The DOA estimation error datasets for these three approaches are derived by comparing their estimation results to the true DOAs and the resulting error statistics are depicted and compared in Fig. \ref{fig:doa_performance}.  

First, Fig. \ref{fig:doa_performance}(a) presents the empirical CDF curves for these three DOA estimation error datasets. It shows that, for this antenna array, since its real manifold after installation severely deviates from the nominal manifold, calibrating it using chamber measurements even slightly deteriorates the DOA estimation performance.
On the contrary, the in-situ estimated array manifold captures the real array responses more precisely by utilizing the post-established in-field measurements.
Specifically, as illustrated by empirical CDF curves in Fig. \ref{fig:doa_performance}(a), through in-situ calibration, a reduction of \(46.7\%\) (from \(3.0^\circ\) to \(1.6^\circ\)) for the \(68\)-th percentile (\(1{\text -}\sigma\)) error and a reduction of \(23.1\%\) (from \(5.2^\circ\) to \(4.0^\circ\)) for the \(90\)-th percentile error are achieved.

Then, to reveal more details of the error statistics, we divide the data into five adjacent sets according to their true DOAs and present the box plots of DOA estimation errors at each set in Fig. \ref{fig:doa_performance}(b). It shows that, the non-calibration and chamber calibration results are obviously biased, while the in-situ calibration errors are all nearly centered at zero. This clearly demonstrates that, array errors offset the DOA estimates, and by in-situ calibration, these offsets are corrected. 

\section{Conclusion}
\label{sec:conclusion}
An in-situ calibration framework and an array manifold estimation algorithm
have been proposed in this work to support high-accuracy 5G positioning.
This framework reduces calibration costs by using off-the-shelf 5G devices and obviating extra hardware and protocol modifications, and improves calibration accuracy by capturing all kinds of in-field array errors, including those induced after installations, in a direction-dependent array error function.
The proposed estimation algorithm fully exploits the bandwidth resources provided by 5G signals and the super-resolution ability of the EM algorithm to resolve the multipaths in the delay domain, whose calibration accuracy is demonstrated to be superior to methods that only utilize the spatial aperture for multipath resolving.
We believe this paper provides a low-cost and scalable solution to calibrate these pervasively installed RRUs in-situ, thereby enabling the 5G network to provide high-precision positioning and sensing services.

Although in this paper we have set the focus on 5G positioning, the proposed calibration scheme can also be applied to other similar wireless positioning systems, such as Wi-Fi or ultra-wideband, to improve their DOA estimation accuracy. Possible future research directions include: (i) investigating the in-situ calibration method for planar arrays to support 3-D positioning, and (ii) studying near-field array calibration to support near-field or mixed far-field and near-field positioning as the far-field condition may not always be guaranteed if the array aperture has been further improved by, for example, the sparse array design or the massive multiple-input multiple-output configuration.

\section*{Nomenclature}
\textbf{Abbreviations}
\begin{IEEEdescription}[\IEEEusemathlabelsep\IEEEsetlabelwidth{\(\mathcal{CN}(\mu, \sigma^2)\)}]
\item [2-D/3-D] Two-/three-dimensional.
\item [3GPP] Third generation partnership project.
\item [5G] Fifth-generation mobile communications technology.
\item [BB] Baseband.
\item [BBU] Baseband unit.
\item [CDF] Cumulative distribution function.
\item [CFR] Channel frequency response.
\item [CIR] Channel impulse response.
\item [CPP] Calibration pilot position.
\item [CSI] Channel state information.
\item [DOA] Direction-of-arrival.
\item [EM] Expectation-maximization.
\item [FFT] Fast Fourier transform.
\item [gNB] Next-generation Node-B.
\item [HPBW] Half-power beamwidth.
\item [IF] Intermediate frequency.
\item [i.i.d.] Independent and identically distributed.
\item [LOS] Line-of-sight.
\item [MLE] Maximum likelihood estimation.
\item [NLOS] Non-line-of-sight
\item [PE] Principal eigenvector.
\item [RRU] Remote radio unit.
\item [TOA] Time-of-arrival.
\item [UL-SRS] Uplink-sounding reference signal.
\item [ULA] Uniform linear array.
\item [UT] User terminal.
\item [VSWR] Voltage standing wave radio.
\end{IEEEdescription}

\textbf{Notations}
\begin{IEEEdescription}[\IEEEusemathlabelsep\IEEEsetlabelwidth{\(\mathcal{CN}(\mu, \sigma^2)\)}]
\item[\(\jmath\)] Imaginary unit (\(\sqrt{-1}\)). 
\item[\((\cdot)^{\mathsf{T}}\)] Transpose operator.
\item[\((\cdot)^{\mathsf{H}}\)] Conjugate transpose operator.
\item[\((\cdot)^*\)] Conjugate operator.
\item[\((\cdot)^{-1}\)] Inverse of a square matrix.
\item[\(|a|\)] Modulus of the complex number \(a\).
\item[\(\angle a\)] Phase (a.k.a. argument) of the complex number \(a\).
\item[\(\|\cdot\|\)] \(\ell_2\)-norm of a vector.
\item[\({[}\mathbf{a}{]}_n\)] \(n\)-th element of vector \(\mathbf{a}\).
\item[\({[}\mathbf{A}{]}_{m,n}\)] Element at \(m\)-th row and \(n\)-th column of matrix \(\mathbf{A}\)
\item[\(\odot\)] Hadamard (element-wise) matrix product
\item[\(\mathcal{CN}(\mu, \sigma^2)\)] Complex Gaussian distribution parameterized by \(\mu\) and \(\sigma^2\)
\item[\(\mathcal{U}(a, b{]}\)] Uniform distribution from \(a\) to \(b\).
\item[\(a(x)\)] Scalar-valued function with the input variable of \(x\).
\item[\(\mathbf{a}(x)\)] Vector-valued function with the input variable of \(x\).
\item[\(x = \mathcal{O}(a)\)] \(\exists k_{1}, k_{2}>0\), such that \(k_{2} \cdot a \leq x \leq k_{1} \cdot a\). 
\end{IEEEdescription}

\balance
\bibliographystyle{IEEEtran}
\bibliography{insitu_calibration_em}

\end{document}